\documentclass{aa}  

\usepackage{graphicx}
\usepackage{txfonts}
\usepackage[]{hyperref}

\hypersetup{
    colorlinks=true,
    citecolor=blue,
    linkcolor=blue,
    filecolor=magenta,      
    urlcolor=cyan,
}

\usepackage{xcolor}

\usepackage{natbib}
\bibpunct{(}{)}{;}{a}{}{,} % to follow the A&A style
\begin{document} 

   \titlerunning{Improved strong lensing modelling of Abell S1063 using the Fundamental Plane}
   \authorrunning{G. Granata et al.}
   \title{Improved strong lensing modelling of galaxy clusters using the Fundamental Plane: Detailed mapping of the baryonic and dark matter mass distribution of Abell S1063}

   \author{G. Granata
          \inst{1},
          A. Mercurio
          \inst{2},
          C. Grillo
          \inst{1,3},
          L. Tortorelli
          \inst{4,5},
          P. Bergamini
          \inst{1,6},
          M. Meneghetti
          \inst{6,7,8},
          P. Rosati
          \inst{9,6}, \\
          G. B. Caminha
          \inst{10},
          \and
          M. Nonino
          \inst{11}
          %\and
         % ????\inst{2}\fnmsep\thanks{Just to show the usage
          %of the elements in the author field}
          }

   \institute{Dipartimento di Fisica, Universit\`a degli Studi di Milano, Via Celoria 16, I-20133 Milano, Italy\\
              \email{giovanni.granata@unimi.it}
            \and
              INAF -- Osservatorio Astronomico di Capodimonte, Via Moiariello 16, I-80131 Napoli, Italy
            \and
              Dark Cosmology Centre, Niels Bohr Institute, University of Copenhagen, Jagtvej 128, DK-2200 Copenhagen, Denmark
            \and
            University Observatory, Faculty of Physics, Ludwig-Maximilian-Universit\"at M\"unchen, Scheinerstr. 1, D-81679 Munich, Germany
            \and
            Institute for Particle Physics and Astrophysics, ETH Zürich, Wolfgang-Pauli-Str. 27, CH-8093 Zürich, Switzerland
            \and
            INAF -- OAS, Osservatorio di Astrofisica e Scienza dello Spazio di Bologna, via Gobetti 93/3, I-40129 Bologna, Italy
            \and
            National Institute for Nuclear Physics, viale Berti Pichat 6/2, I-40127 Bologna, Italy
            \and
            Division of Physics, Mathematics, \& Astronomy, California Institute of Technology, Pasadena, CA 91125, USA
            \and
            Dipartimento di Fisica e Scienze della Terra, Universit\`a degli studi di Ferrara, via Saragat 1, I-44122 Ferrara, Italy
            \and
            Max-Planck-Institut f\"ur Astrophysik, Karl-Schwarzschild-Str. 1, D-85748 Garching, Germany
            \and
            INAF -- Osservatorio Astronomico di Trieste, via G. B. Tiepolo 11, I-34131 Trieste, Italy
           }

   \date{January 25, 2022}

  \abstract
   {} %leave it empty if necessary  
   {From accurate photometric and spectroscopic information, we build the Fundamental Plane (FP) relation for the early-type galaxies of the cluster Abell S1063. We use this relation to develop an improved strong lensing model of the cluster, and we decompose the cluster's cumulative projected total mass profile into its stellar, hot gas, and dark matter mass components. We compare our results with the predictions of cosmological simulations.}
   {We calibrate the FP using Hubble Frontier Fields photometry and data from the Multi Unit Spectroscopic Explorer on the Very Large Telescope. The FP allows us to determine the velocity dispersions of all $222$ cluster members included in the model from their measured structural parameters. As for their truncation radii, we test a proportionality relation with the observed half-light radii. Fixing the mass contribution of the hot gas component from X-ray data, the mass density distributions of the diffuse dark matter haloes are optimised by comparing the observed and model-predicted positions of $55$ multiple images of $20$ background sources distributed over the redshift range $0.73-6.11$. We determine the uncertainties on the model parameters with Monte Carlo Markov chains.}
   {We find that the most accurate predictions of the positions of the multiple images are obtained when the truncation radii of the member galaxies are approximately $2.3$ times their effective radii. Compared to earlier work on the same cluster, our model allows for the inclusion of some scatter on the relation between the total mass and the velocity dispersion of the cluster members. We notice a lower statistical uncertainty on the value of some model parameters. For instance, the main dark matter halo of the cluster has a core radius of $86 \pm 2 \, \mathrm{kpc}$: the uncertainty on this value decreases by more than 30\% with respect to previous work. Taking advantage of a new estimate of the stellar mass of all cluster members from the HST multi-band data, we measure the cumulative two-dimensional mass profiles out to a radius of $350 \, \mathrm{kpc}$ for all baryonic and dark matter components of the cluster. At the outermost radius of $350 \, \mathrm{kpc}$, we obtain a baryon fraction of $0.147 \pm 0.002$. We study the stellar-to-total mass fraction of the high-mass cluster members in our model, finding good agreement with the observations of wide galaxy surveys and some disagreement with the predictions of halo occupation distribution studies based on $N$-body simulations. Finally, we compare the features of the sub-haloes as described by our model with those predicted by high-resolution hydrodynamical simulations. We obtain compatible results in terms of the stellar over total mass fraction. On the other hand, we report some discrepancies both in terms of the maximum circular velocity, which is an indication of the halo compactness, and the sub-halo total mass function in the central cluster regions.}
   {}


   \maketitle
%
%-------------------------------------------------------------------
\section{Introduction}\label{s1}

   The concordance, or \textrm{$\Lambda$} cold dark matter (CDM), cosmological model, dominated by CDM and with a cosmological constant $\Lambda$, allows for specific predictions about the physical properties of dark matter (DM) haloes at different scales. Cold dark matter drives the hierarchical growth of structures in the Universe: the most massive DM haloes form from mergers of smaller objects and by accretion \citep[e.g.][]{tormen97,moore99}. Many of the less massive haloes are therefore found orbiting around the largest structures: they are usually referred to as sub-haloes or the substructures of a main halo. $N$-body cosmological simulations can be used to predict the number of sub-haloes around a halo of a given mass. They also show that DM haloes of any mass have a similar mass density profile, with a central cusp, usually described using the Navarro-Frenk-White \citep[][]{navarro96,navarro97} or Einasto \citep[][]{einasto65} profiles \citep[e.g.][]{wang20}. Any significant discrepancy between these predictions and the observations may imply that the formation of structures does not proceed as forecast by the \textrm{$\Lambda$}CDM model.
   
   Galaxy clusters are especially well suited for testing the predictions of cosmological simulations. Around $85-90\%$ of their mass is made up of DM located in cluster-scale haloes and galaxy-scale haloes or sub-haloes. The remaining $10-15\%$ is composed of baryons: hot X-ray-emitting plasma and stars, which represent only approximately $1\%$ of the total mass \citep[][]{annunziatella17,sartoris20}. The entire baryonic mass budget of clusters can be measured from photometric and spectroscopic observations in the near-IR, visible, and X-ray bands. Thus, if one determines the total mass distribution, one can disentangle that of the DM haloes and compare it with simulations \citep[e.g.][]{diemand11}.
   
   In the past few years, gravitational lensing has become one of the most effective techniques for studying the total mass distribution of clusters \citep[e.g.][]{natarajan97} as it does not discriminate between baryons and DM. Strong lensing (SL) is especially effective in the central regions, or cores, of massive clusters (i.e. within approximately $1$ arcmin, or approximately $350$ kpc at $z=0.5$), where several multiple images of background sources are usually found.
   In these regions, two important discrepancies between the predictions of $N$-body simulations and lensing models have emerged. First, several SL models suggest that the main DM haloes of clusters might have central flat cores \citep[][]{sand04}, extended up to around 100 kpc \citep[][hereafter B19]{grillo15,annunziatella17,bergamini19}. This result seems to be confirmed on a homogenous and unbiased sample of massive lens galaxy clusters \citep[][]{postman12}. A second discrepancy appears when comparing the amount and distribution of DM substructures (or sub-haloes). As found in \citet{grillo15}, \citet{munari16}, and \citet{bonamigo18}, $N$-body simulations underestimate the number of massive substructures near the cores of massive clusters. Furthermore, the number of predicted galaxy-scale SL events seems to imply that the mass distributions of these substructures are more concentrated than suggested by simulations \citep[][hereafter M20]{meneghetti20}. 
   
   Several Hubble Space Telescope (HST) programmes have recently been dedicated to the observation of SL phenomena caused by massive clusters. For instance, the Cluster Lensing And Supernova survey with Hubble \citep[CLASH;][]{postman12}, the Hubble Frontier Fields \citep[HFF;][]{lotz17} campaign, and the Reionization Lensing Cluster Survey \citep[RELICS;][]{coe19} obtained multi-band HST images with an unprecedented depth for 25, 6, and 41 clusters, respectively. Moreover, spectroscopic campaigns such as CLASH-VLT \citep[Dark Matter Mass Distributions of Hubble Treasury Clusters and the Foundations of $\Lambda$ CDM Structure Formation Models, PI: P. Rosati; see][]{rosati14}, complemented with data from the new integral field Multi Unit Spectroscopic Explorer \citep[MUSE;][]{bacon10} on the Very Large Telescope (VLT), have permitted secure identifications of cluster members and precise redshift measurements for hundreds of multiply lensed sources.
   These data have allowed for the construction of detailed SL models, whose accuracy can be assessed by considering the distance between the observed positions of multiple images and those predicted by a model. The root mean square (rms) of this discrepancy can be as low as $0.3''$ \citep[e.g. B19,][]{caminha19}; however, this is still higher than the observational uncertainty on the position of the images, which can be less than a tenth of an arcsecond. This might suggest that the total mass modelling of galaxy clusters needs to be further improved \citep[see][]{meneghetti17,remolina18,raney20}. 
   
   Strong lensing observations permit a very effective determination of the total mass enclosed within the multiple images of a single background source, but the same amount of mass can be obtained with relatively different mass distribution models \citep[e.g.][]{limousin16,ghosh21}. This implies that the parameters that define a model often have a certain degree of degeneracy between them, which corresponds to a possible transfer of mass between the various components. 
   Specifically, galaxy clusters are modelled with extended, cluster-scale mass clumps, to describe the main DM haloes \citep[][]{natarajan97} and the intra-cluster medium (ICM), and with smaller haloes, to represent the member galaxies \citep[][]{delucia04}. This last component is generally parametrised in SL models with spherical, truncated, isothermal profiles, defined only by their values of effective velocity dispersion and truncation radius, in order to reduce the number of optimised free parameters. Furthermore, their values are usually linked to the luminosity of each cluster member with two power-law scaling relations \citep[e.g.][]{richard14,grillo15,monna17}. This approach is equivalent to using a Faber-Jackson scaling law \citep[][]{faber76} to determine the value of the velocity dispersion of an elliptical galaxy.
   
   Information on the kinematics of galaxies can significantly reduce the degeneracy between the parameters of an SL model. Observations of galaxy-scale SL events by the Sloan Lens ACS (SLACS) survey allowed for a precise determination of the slope of the total mass density profile of massive lens galaxies \citep[][]{koopmans06}. In recent years, velocity dispersion measurements have also become available for several members in some galaxy clusters. \citet{monna15} used spectrography from the Hectospec fibre spectrograph at the MMT and early CLASH data for 21 members of the cluster Abell 383, showing a significant reduction in the degeneracy between the SL model parameters. 
   B19 refined and significantly extended this method to three CLASH clusters: MACS J1206.2$-$0847, MACS J0416.1$-$2403, and Abell S1063. The integral field data provided by MUSE permitted velocity dispersion measurements to be obtained for 40$-$60 galaxies for each cluster. These data have been exploited to estimate the values of the slope and normalisation of the Faber-Jackson relation for those members. This has, in turn, been used as a prior to determine the exponent of the power-law relation that binds the total mass of sub-haloes to their luminosity.
   The new data and procedure have allowed for a more accurate mass assignment to the diffuse and sub-halo components, reducing the degeneracy between their parameters. This is crucial for both inferring the DM mass density profile in the central regions of a cluster and precisely determining its sub-halo mass function.
   
   This technique has enabled a first step towards a more accurate mass modelling of the cluster members, but a single power-law relation between their total mass and luminosity may be inaccurate on the very large total mass range of member galaxies ($10^9-10^{13} M_{\odot}$). In order to improve on the adopted scaling relations for members, in this work we consider the so-called Fundamental Plane (FP) relation \citep[][]{dressler87,djorgovski87,bender92}, which has a lower scatter than the Faber-Jackson law. We focus on one of the three clusters considered by B19, namely Abell S1063 (AS1063). We chose it due to its rather regular morphology and because we have at our disposal line-of-sight velocity dispersion measurements for a sizable subset of cluster members, for which we also have effective radius and magnitude estimates. We can therefore use these galaxies to obtain the best-fit values of the parameters that define the FP. 
   
   The paper is organised as follows. In Sect. \ref{s2} we give details on our photometric and spectroscopic data and on how they were analysed and reduced for our objectives. In Sect. \ref{s3} we describe our SL model built with the FP, while in Sect. \ref{s4} we present the analysis and discussion of the results of the optimisation and the comparison with the previous model from B19. In Sect. \ref{s5} we show the cumulative, projected mass profiles for all the baryonic and DM components of our model. In Sect. \ref{s6} we compare our results with the predictions of stellar-to-halo mass relations derived from $N$-body simulations. In Sect. \ref{s7} we compare the stellar-to-total mass fraction and the compactness of the sub-haloes in our model with those suggested by recent high-resolution hydrodynamical simulations. Finally, in Sect. \ref{s8} we summarise our most important conclusions. 
   
   In this work, we use a flat $\mathrm{\Lambda}$CDM cosmology with $\Omega_\mathrm{m}=0.3$ and $H_0 = 70$ $\mathrm{km \, s^{-1} \, Mpc^{-1}}$, in which $1''$ corresponds to a scale of 4.92 kpc at $z=0.348$, the redshift of AS1063. All magnitudes are expressed in the AB system.
   
%--------------------------------------------------------------------
\section{Abell S1063: Data and analysis}\label{s2}

   Abell S1063, first identified in \citet{abell89}, is a massive galaxy cluster at redshift $z=0.348$. A virial mass value of $(2.17 \pm 0.06) \times 10^{15} \, M_\odot$ was estimated by \citet{sartoris20}. Despite its regular shape, it appears to have recently undergone an off-axis merger, as inferred by \citet{gomez12} and \citet{rahaman21} from X-ray data, and by \citet{mercurio21} from the dynamics of the members. Its high mass and redshift make it a very effective gravitational lens, which allows us to use the observed position of 55 spectroscopically confirmed multiple images from 20 background sources, distributed over a wide range of redshifts \citep[$z=0.73-6.11$; see][]{caminha16}, to build an SL model. This high number of spectrocopically confirmed sources is crucial for an accurate SL model. Indeed, the well-known degeneracy between the total mass of a lens and the redshift of a multiply imaged source can be broken when several multiply imaged sources in a wide redshift range are available and their redshift are measured.
   
   The cluster is also a very bright X-ray source ($L_X \approx 2.5 \times10^{45} \, \mathrm{erg \, s^{-1}}$ in the $0.1-2.4 \, \mathrm{keV}$ band), due to the emission of the hot ($T \approx 12 \, \mathrm{keV}$) ICM \citep[][]{rahaman21}: it was identified as RXJ2248.7$-$4431 in the ROSAT All-Sky Survey \citep[][]{degrandi99, guzzo99}.
   
   \subsection{Photometric data}
   
   Abell S1063 has been included in the two main photometric campaigns by the HST that, in the last decade, targeted SL phenomena in galaxy clusters. Firstly, it was one of the 25 clusters included in CLASH. Both the Advanced Camera for Surveys (ACS) and the Wide Field Camera 3 (WFC3) instruments were used to observe the clusters with 16 different broadband filters, from the near-UV to the near-IR. An even deeper view on AS1063 was obtained thanks to the HFF programme, which was assigned 840 HST orbits to target a selected sample of only six clusters, in seven different bands (HST/ACS F435W, F606W, and F814W and HST/WFC3 F105W, F125W, F140W, and F160W). Following \citet{tortorelli18}, we used the HFF photometric data in the F814W band, which corresponds to a rest-frame R-band for the cluster redshift, as it allows for a good signal-to-noise ratio down to magnitudes of around 28 and it has a lower full width at half maximum of the point spread function when compared to the near-IR bands.
   
      \begin{figure*}
   \centering
   \includegraphics[width=17cm]{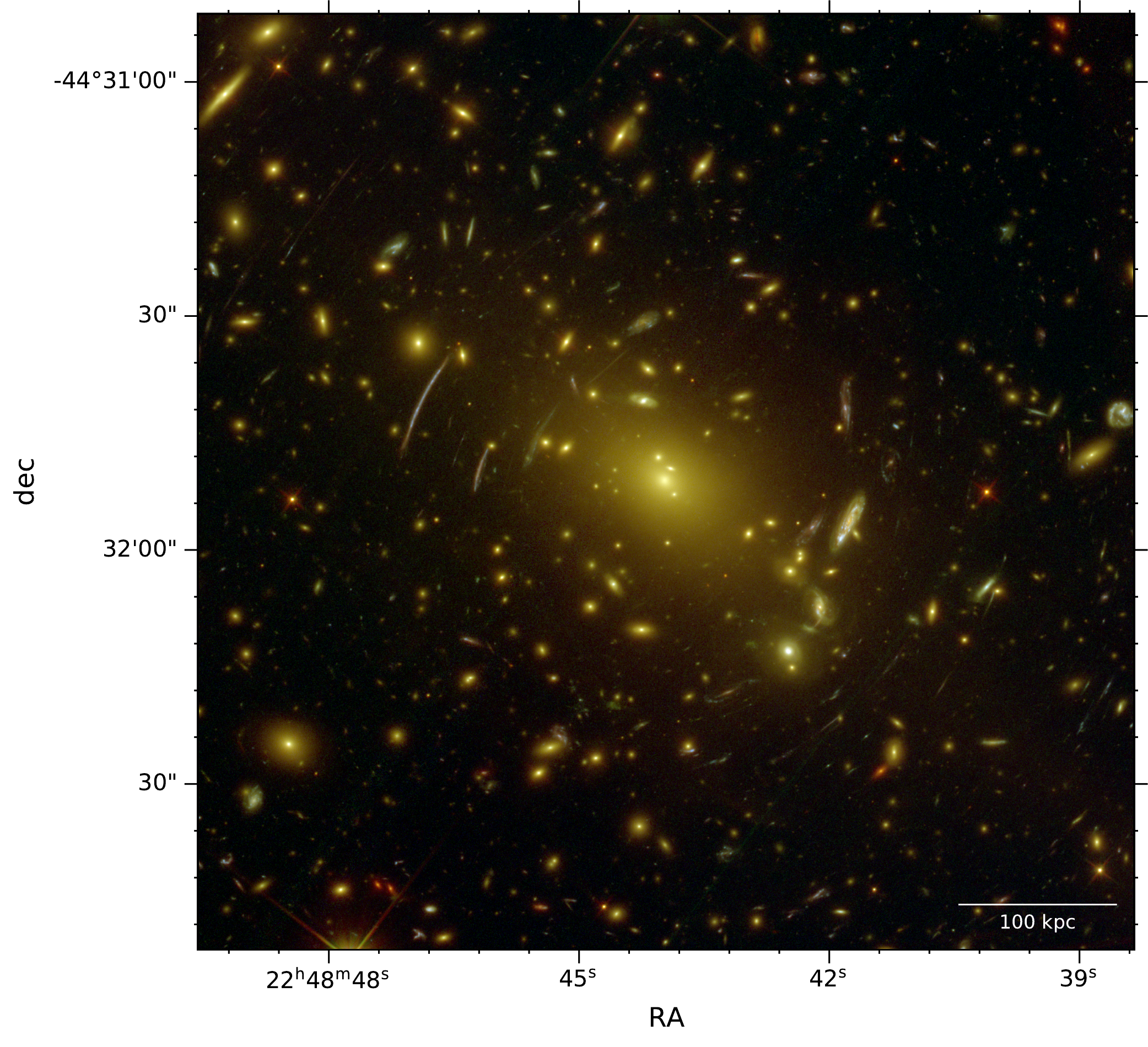}
      \caption{Colour composite image of the core of AS1063 created from HFF photometry. Blue, green, and red channels are obtained by combining the filters F435W, F606W + F814W, and F105W + F125W + F140W + F160W, respectively.}
         \label{f814hst}
   \end{figure*}
   
   In order to obtain the parameters of the FP and consequently derive the values of the velocity dispersions of the cluster members, we need robust measurements of their magnitudes and effective radii. We adopt the prescriptions developed in \citet{tortorelli18} and further enhanced in an upcoming work from the same authors to analyse the 222 galaxies included in our model. The galaxy surface brightness profile fitting consists of iterative steps, where we consecutively increase the size of the fitted images. We started by measuring the photometric properties with \texttt{SExtractor} \citep[][]{bertin96}. These properties were used as initial guesses for the more accurate galaxy light profile analysis with \texttt{GALFIT} \citep[][]{peng10}. We cut stamps around each galaxy, fitting its light profile and those of the closest galaxies. Then, the best-fitting parameter values were used as starting points for the following iterations, where we simultaneously fitted all galaxies falling in the nine regions we separated our image into. Finally, the parameters obtained in the previous step were used as a starting point for the final simultaneous light profile fitting of all galaxies in the whole image.
   
   The surface brightness profiles of the galaxies were fitted with \texttt{GALFIT} using single S\'ersic light profiles with seven free parameters (the two centre coordinates, major-to-minor axis ratio, position angle, major-axis effective radius, S\'ersic index, and magnitude). For every galaxy, \texttt{GALFIT} was provided with different combinations of point spread functions, background value estimates, and noise images, as described in \citet{tortorelli18}. This allowed us to average out the contribution of systematic effects in the surface brightness fit.

   The spectral energy distribution obtained from the multi-band HST photometry also permits us to infer the stellar mass of the member galaxies, using composite stellar population models. Further details are provided in \citet{mercurio21}, where the authors used the HST photometry plus the MUSE spectra to fit the spectral energy distributions of the members with a delayed exponential star formation history, solar metallicity, and a Salpeter \citep[][]{salpeter55} stellar initial mass function (IMF). The presence of dust was taken into account following \cite{calzetti00}. %There is a known degeneracy between age and metallicity of a stellar population, which can be reduced including spectroscopic data in the fit, in order to determine the age of stars more accurately.
   
   However, MUSE data are only available for a fraction of our sample. For those galaxies we compare their stellar mass values obtained including information from spectra with those derived without. We notice only a small systematic offset between the two estimates. Therefore, we chose to consider, for all the members, the stellar masses obtained by using only the photometric data so as to have a uniform sample. From this comparison, we also fixed a conservative error of $0.22$ on the decimal logarithmic value of the stellar mass.
   
   \subsection{Spectroscopic data}
   
   Abell S1063 was targeted by several spectroscopic programmes. It was one of the 13 clusters included in the CLASH-VLT  survey. A total of 16 masks were used on a field of approximately $25$ arcmin across, allowing for the measurement of 3607 reliable redshifts, 1109 of which identify cluster members \citep[][]{mercurio21}.
   
   The integral field spectrograph MUSE has complemented the Visible Multi Object Spectrograph (VIMOS) observations in cluster cores. Abell S1063 was observed with two different pointings, centred in the NE and SW regions of the cluster \citep[][]{karman15,caminha16,karman17}, with an exposure time of $3.1$ and $4.8$ hours and a seeing of $1.1''$ and $0.9''$, respectively. MUSE provided a total of 175 additional reliable redshifts, 104 of which refer to cluster members. Following \citet{caminha16}, we identify as spectroscopic cluster members those galaxies whose redshift falls in the range $0.336-0.362$, corresponding to a rest-frame velocity range of $\pm 3000 \, \mathrm{km} \, \mathrm{s}^{-1}$. Moreover, multi-band CLASH data permit photometric members to be identified with the method presented in \citet{grillo15}.
   
   The MUSE data are deep enough for the measurement of the line-of-sight stellar velocity dispersion of a considerable fraction of the cluster members (80 out of 222). As in B19, we used the \texttt{pPXF} software (penalised pixel-fitting) by \citet{cappellari04} for the measurement, but instead of using a fixed circular aperture of $0.8''$ for all members, we chose a wider one, of $2''$, and weigh each pixel with its corresponding surface brightness value: this choice has been found to provide velocity dispersion values equivalent to the central ones. The aperture was reduced to $0.5''$ for a member whose velocity dispersion measurement was strongly influenced by the effects of the diffuse light of the brightest cluster galaxy (BCG). \texttt{pPXF} estimates the line-of-sight-velocity-distribution (LOSVD) parameters by comparing the observed spectra with a set of stellar templates from the MILES library \citep[][]{vazdekis10} convolved with a LOSVD. The code minimises a $\chi^2$ function between the measured spectrum and a model.
   
\section{Building the strong lensing model}\label{s3}
   
   We built our SL model using that from B19 as a basis and taking advantage of the information provided by the FP relation for a more accurate determination of the total mass of the cluster members. We then compared our model with the B19 model to understand how the new modelling choices regarding the mass of the sub-haloes influence the reconstruction of the diffuse mass component of the cluster.
   
   As in B19, we started from the same 55 spectroscopically confirmed multiple images produced by 20 sources presented in \citet{caminha16}. Likewise, we modelled the various lens mass components with dual pseudo-isothermal elliptical (dPIE) mass density profiles \citep[][]{limousin05,eliasdottir07}, which are the ellipsoidal generalisation of a truncated isothermal sphere with a central core. As in B19, we used the publicly available SL code \texttt{LensTool} \citep[][]{kneib96,jullo07,jullo09} in order to find the best-fit values of the parameters of our model. The dPIE profile is defined by seven free parameters: two for the centre position, the ellipticity $e$ and its position angle $\theta_e$, the central velocity dispersion $\sigma_0$, and the core and truncation radii $r_c$ and $r_t$, respectively. The projected mass density of a spherical dPIE as a function of the two-dimensional distance from the centre $R$ is \citep[][]{eliasdottir07}
   \begin{equation}
       \Sigma (R) = \frac{\sigma_0^2}{2G} \frac{r_t}{r_t-r_c} \left(\frac{1}{\sqrt{r_c^2+R^2}}-\frac{1}{\sqrt{r_t^2+R^2}} \right).
   \end{equation}
    \texttt{LensTool} actually parametrises the dPIE profile with a fiducial velocity dispersion that is related to the central velocity dispersion by $\sigma_\mathrm{LT} = \sqrt{2/3} \sigma_0$ \citep[see][and Appendix C in B19]{eliasdottir07,agnello14}. As the projected mass of a sub-halo within a given aperture depends on its truncation radius and central velocity dispersion, the same enclosed mass can be obtained with different combinations of values for these two parameters. As shown by B19, the introduction of a kinematic prior significantly reduces this degeneracy, allowing one to obtain total mass values that are consistent with the kinematic data.
   The diffuse, cluster-scale, DM haloes were represented with an elliptical dPIE, close to the BCG position, and a spherical dPIE with vanishing core radius, whose position was left free to vary on an area of $150'' \times 120''$ centred in the NE region of the cluster. This second halo is necessary to reproduce some arcs in the NE region and its presence is compatible with the post-merger scenario of the cluster. As in B19, the hot gas mass component is completely fixed by the Chandra X-ray data, reduced in \citet{bonamigo18}. This component was modelled with three dPIE mass profiles, whose parameters were not optimised. The position of each multiple image and source is identified by two coordinates. As the positions of the sources are not known, 55 multiple images and 20 sources correspond to, respectively, 110 observables and 40 free parameters. The two cluster-scale DM clumps have nine further parameters: the values of the coordinates of the centre and of the velocity dispersions were left free for both haloes, while those of the core radius, ellipticity and orientation angle were only optimised for the main halo. There is therefore a total of 49 free parameters, leading to 61 degrees of freedom.
   
   As anticipated, the main difference with the model obtained in B19 is in the way we assign mass to the cluster members. They were modelled as spherical dPIE profiles with a negligible core radius, so that their total mass is entirely fixed by their values of velocity dispersion and truncation radius. Their three-dimensional total mass density $\rho(r)$ scales as $r^{-2}$ inside the truncation radius, and $r^{-4}$ outside. Within a sphere with radius equal to the truncation radius, $50\%$ of the total mass is enclosed, while approximately $90\%$ of it is enclosed within a six times larger radius. In order to estimate the velocity dispersion of member galaxies from their measured effective radius and surface brightness, B19 calibrated, for a subset of the members, the Faber-Jackson relation between the observed values of luminosity and line-of-sight velocity dispersion. The resulting relation was used to impose a prior on the exponent of an optimised power-law scaling relation between the two observables, valid for all members. With this approach, a fixed law, with no scatter, is imposed to describe members in a wide mass range. Furthermore, the total mass of members only depends on a single measured quantity, their total luminosity, and might therefore be more sensitive to observational offsets or errors. 
   
   The Faber-Jackson law has a scatter of around $0.1$ on the logarithmic value of the velocity dispersion \citep[][]{davies83}, and the residuals show a significant correlation with galaxy size \citep[e.g.][]{fraixburnet10,nigochenetro11}. This indicates that there is a tighter relation, which includes the effective radius of galaxies, known as the FP \citep[][]{dressler87,djorgovski87,bender92}, of which the Faber-Jackson law is just a projection. The fact that the FP deviates from the expectations for homologous galaxies in virial equilibrium can be explained with the hypothesis of a variation in the total mass-to-light ratio of galaxies with their total mass \citep[][]{faber87}, or of a non-homologous structure of the galaxies along the FP \citep[][]{djorgovski95,hjorth95,desmond17}. 
   
   As anticipated, in our model we used the FP relation to completely fix the value of the velocity dispersion of the members from their measured effective radius and total magnitude values. This procedure has several advantages. First, the FP has a lower scatter of $0.07$ on the logarithmic value of the line-of-sight velocity dispersion \citep[][]{jorgensen96}. The best-fit parameters of the relation were determined with an optimisation performed on a selected sample of members, to avoid the effects of systematics affecting the observations. Furthermore, the value of the velocity dispersion of a member depends on two observed quantities, instead of one. This method is therefore less sensitive to possible measurement errors on a specific quantity.
   Late-type galaxies were included with early-type galaxies and modelled in the same way. We are aware that this is an approximation, but, as clearly visible in Fig. 1, late-type galaxies are a very small fraction of the members located in the cluster central regions. Moreover, the most massive cluster members, providing the most important contribution to the SL model, are early-type galaxies. 
   
   As for the truncation radii of the members, we chose them to be proportional to their observed half-light radii. This ansatz is suggested by the abundance matching results of \citet{kravtsov13}, who found that the virial and half-light radii of galaxies are approximately proportional, as previously suggested by some theoretical studies \citep[e.g.][]{mo98}. The ratio between the two mostly depends on the angular momentum acquired from tidal torque by the halo during its formation. The relation was observed for galaxies within a very wide range of stellar masses and morphologies, including elliptical galaxies in the Virgo cluster. In our case, rather than the virial radii of members, we are interested in estimating their truncation radii, which are significantly smaller. Galaxies in the cluster core might have a generally self-similar total mass profile, leading to an approximately constant ratio between truncation and virial radii, as a consequence of similar evolution processes. This leads to an approximately proportional relation between effective and truncation radius, which is the ansatz we used in this paper. Compared to previously adopted power-law relations between the truncation radius and the total luminosity of a member, our assumption relates in a more direct way two spatial scales of the members. We calibrated the proportionality relation by testing several values for the truncation-to-effective radius ratio, optimising all the free parameters describing the other mass components of the cluster each time. Finally, we chose the ratio value that allows for the lowest rms value of the discrepancy between the observed and model-predicted positions of the multiple images considered in the optimisation of the model. 
    
\subsection{The Fundamental Plane of Abell S1063}

    The FP is an empirical relation between the logarithm of the effective radius (i.e. the bi-dimensional radius inside which half of the total luminosity is emitted) $R_e$, the average surface brightness within it, $\mathrm{SB}_e$, and the logarithm of the central stellar velocity dispersion $\sigma_0$. The FP can be written as
    \begin{equation} \label{fp}
    \log R_e = \alpha \log \sigma_0 + \beta \mathrm{SB}_e + \gamma.
    \end{equation}
    We consider $R_e$ as measured in kpc, and $\sigma_0$ in $\mathrm{km \, s^{-1}}$, whereas $\mathrm{SB}_e$ is defined as $(1 \, \mathrm{mag \, arcsec^{-2}})\left(\frac{m}{1 \, \mathrm{mag}}+2.5\log \left(\frac{2\pi R_e^2}{1 \, \mathrm{kpc^2}}\right)\right)$, where $m$ is the total magnitude measured using \texttt{GALFIT}.

    As anticipated, MUSE observations and their subsequent analysis provide us with line-of-sight central stellar velocity dispersions, $\sigma_0$, for 80 out of the 222 member galaxies included in the model. As the FP is actually only valid for early-type galaxies, we need to ensure that the values of its parameters are not influenced by galaxies to which the FP relation does not strictly apply. There are different ways of selecting early-type galaxies, based on their morphologies, colours, and stellar populations. In \citet{tortorelli18}, it has been shown that early-type galaxies selected via a S\'ersic index value $n>2.5$, elliptical morphology and a passive stellar population constitute a coeval population. Therefore, we selected early-type galaxies as those galaxies with S\'ersic index $n>2.5$. Furthermore, we abide by the criteria introduced in B19, based on the robustness of the velocity dispersion measurements by \texttt{pPXF} on simulated spectra, thus including only those with a spectrum signal-to-noise ratio greater than 10 and $\sigma_0>80\, \mathrm{km \, s^{-1}}$. We also excluded two further galaxies: one is in the diffuse light of the BCG, and the second is very close to another member, which influences the velocity dispersion measurement. This leaves us with a total of 30 spectroscopically confirmed cluster members, out of the 80 for which we had a measurement of the stellar velocity dispersion, which we used as a basis to estimate the parameters of the FP relation. 

    To obtain the FP parameter values, we performed an optimisation using the code \texttt{ltsfit}, developed by \citet{cappellari13} for robust fits of lines or planes. A linear regression algorithm is used on the data to derive a first value for the parameters. A $\sigma$ dispersion around the plane is defined as the range in which 68\% of the values fall. The points that lie at more than $3\sigma$ from the plane determined by the first regression are clipped and the code repeats the procedure iteratively with the remaining points, until no new member is discarded and the final parameter values are obtained. As for the errors on the observables, we used the propagated experimental uncertainty. \citet{tortorelli18} find, for the same cluster, an average uncertainty of $0.09$ $\mathrm{kpc}$ on $R_e$ and of $0.19$ $\mathrm{mag \, arcsec^{-2}}$ on $\mathrm{SB}_e$ for galaxies with $m_\mathrm{F814W} \le 22.5$, which are significantly higher than what we obtain through the error propagation for several members. We chose $0.09$ $\mathrm{kpc}$ and $0.19$ $\mathrm{mag \, arcsec^{-2}}$ as lower limits for those uncertainties. 
    We list the best-fit values of $\alpha$, $\beta$, and $\gamma$ obtained with \texttt{ltsfit} in Table \ref{FPTAB}. The values of $\alpha$ and $\beta$, which do not depend on redshift, are compatible with those reported in the literature \citep[e.g.][]{barr06}.

   \begin{table}
\caption{Best-fit values and $1\sigma$ errors of the parameters of the FP.}             % title of Table
\label{FPTAB}      % is used to refer this table in the text
\centering                          % used for centering table
\begin{tabular}{c c}        % centered columns (4 columns)
\hline \noalign{\smallskip}                % inserts double horizontal lines
Parameter & Value \\    % table heading 
\noalign{\smallskip} \hline  \noalign{\smallskip}                      % inserts single horizontal line
   $\alpha$ & $0.99 \pm 0.17$ \\      % inserting body of the table
   $\beta$ & $0.323 \pm 0.029$ \\
   $\gamma$ & $-10.3 \pm        1.0$ \\
 \noalign{\smallskip} \hline                                   %inserts single line
\end{tabular}
\end{table}
   
    \begin{figure}
   \centering
   \includegraphics[width=9cm]{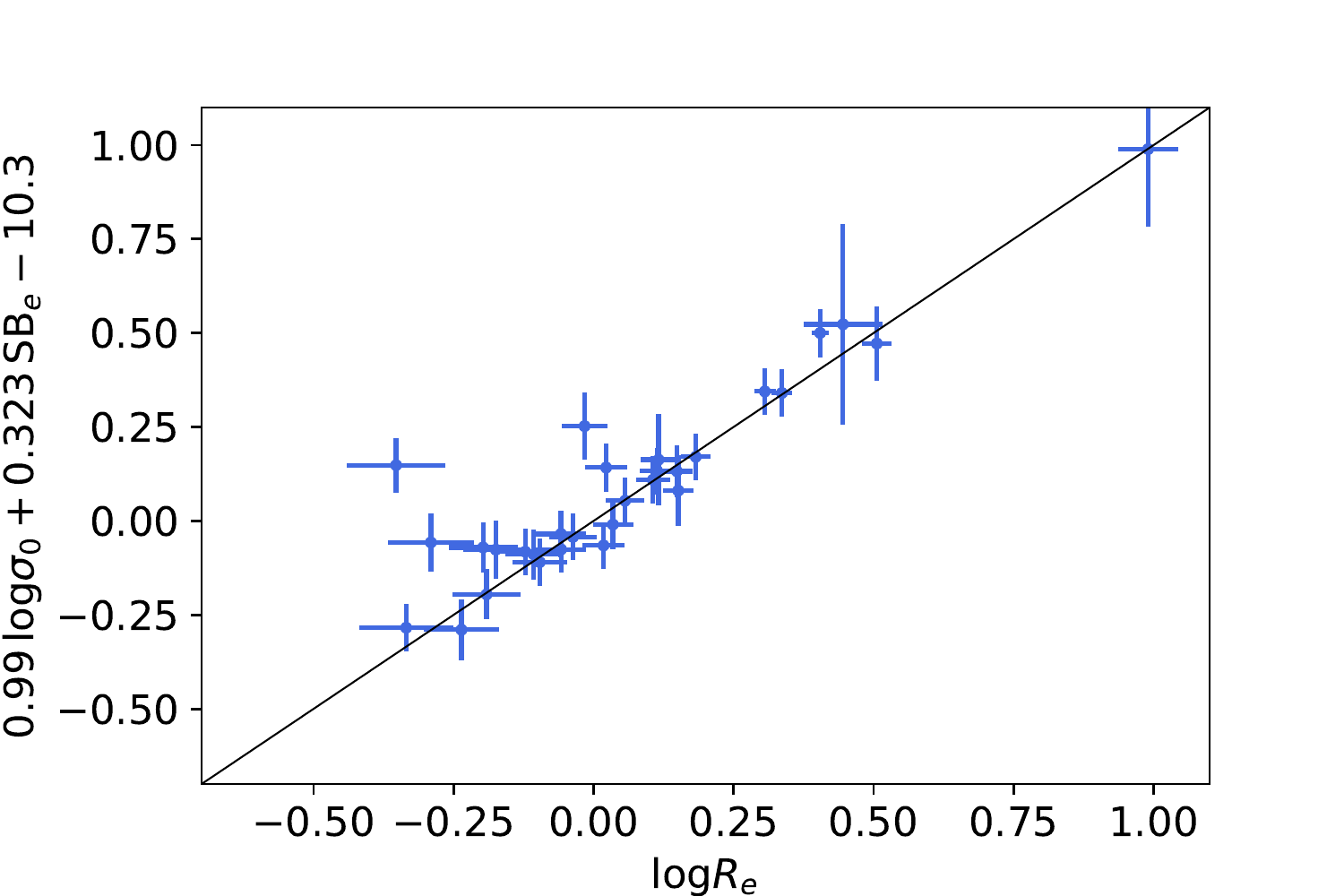}
      \caption{Distribution around the FP of the $30$ members selected to determine its parameters, with the $1\sigma$ uncertainties obtained through the error propagation. For the physical quantities, we use the units listed in the main text.}
         \label{FigVibStab}
   \end{figure}
    
    As is clear from Fig. \ref{FigVibStab}, there are very few outliers around the FP: their photometric and spectroscopic analyses have been checked and confirmed. They do not however influence the values of the FP parameters, as they are clipped during the optimisation procedure. To evaluate the accuracy of the calibrated FP at predicting cluster member central velocity dispersions, we can compare the measured values of $\sigma_0$ with those obtained through the relation for the $30$ members used to derive the parameters. We call these values $\sigma_{0,\mathrm{FP}}$. The median value of the ratio $\frac{\sigma_{0,\mathrm{FP}}}{\sigma_0}$ is $0.98$, with a standard deviation of $0.20$, corresponding to a $\chi_{\mathrm{red}}^2=\sum_{i=1}^{30} \frac{(\sigma_{0,\mathrm{FP} \, i}-\sigma_{0 \, i})^2}{30\sigma_{0 \, i}^2}=0.05$, denoting a low scatter around the plane, as shown by Fig. \ref{FigVibStab}. Furthermore, we do not notice any systematic offsets between the observed values of $\sigma_0$ and those predicted by the FP.
    
\subsection{The best-fit model}
    
    From the best-fit FP parameters, we can compute the value of the central stellar velocity dispersion for all members. As anticipated, we take the value of the truncation radius of the members as a multiple of their measured effective radius, testing several proportionality constant values and re-optimising the values of all the parameters describing the diffuse DM components each time.
    
    To determine the values of the parameters of the best-fit model, we used the $\chi^2$ minimisation implemented in \texttt{LensTool}. The $\chi^2$ function is defined as
    \begin{equation} \label{chi2eq}
    \chi^2 (\boldsymbol{\theta}) = \sum^{N_{\mathrm{fam}}}_{j=1}\sum^{N^j_{\mathrm{img}}}_{i=1} \left( \frac{\lVert \mathbf{x}_{\mathrm{obs}\,i,j} - \mathbf{x}_{\mathrm{pred}\,i,j}(\boldsymbol{\theta})\lVert}{\sigma_{x \, i,j}} \right)^2,
    \end{equation}
    where $N_{\mathrm{fam}}$ is the number of multiple image families considered to evaluate the precision of our model and $N^j_{\mathrm{img}}$ is the number of multiple images that compose the $j$-th family. $\mathbf{x}_{\mathrm{obs} \, i,j}$ and $\sigma_{x \, i,j}$ are, respectively, the observed position of the $i$-th image of the $j$-th family and its uncertainty. Finally, $\mathbf{x}_{\mathrm{pred}\,i,j}(\boldsymbol{\theta})$ is the position of the same image as predicted by the model defined by the set of parameter values $\boldsymbol{\theta}$. As in \citet{bonamigo18} and B19, we assume an error of $0.5''$ on the HST multiple images and of $1''$ for the multiple images only found in the MUSE data.
    
\begin{figure}
  \centering
    \includegraphics[width=9cm]{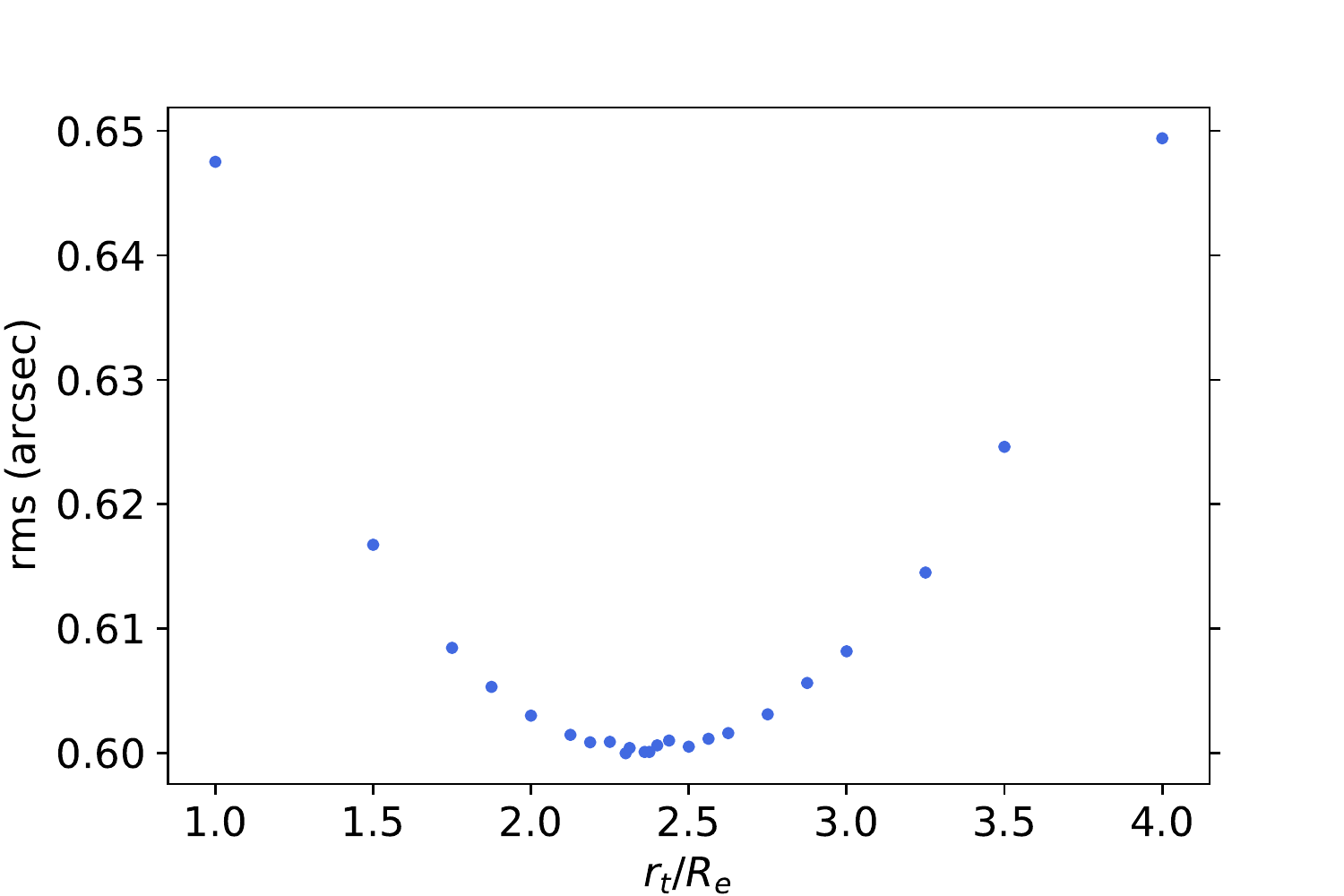}
  \caption{Values of the rms difference between the observed and model-predicted positions of the 55 multiple images used in our model as a function of the ratio between the truncation and effective radii of the cluster members.}
  \label{chi2}
  \end{figure}

    We find that the lowest values of the $\chi^2$ are obtained when the proportionality constant varies between 2.2 and 2.5, as clear from Fig. \ref{chi2}. In particular, we obtain the absolute lowest value of $\chi^2$ for $r_t=2.3R_e$. In this case, we find that the minimum $\chi^2$ has a value of $84.92$, with 61 degrees of freedom, which corresponds to a rms value of $0.60''$. 
    
   \begin{table}
\caption{Best-fit values and $1\sigma$ errors of the parameters of our SL model, compared with those obtained in B19.}             % title of Table
\label{bestfit}      % is used to refer this table in the text
\centering                          % used for centering table
\begin{tabular}{c c c}        % centered columns (4 columns)
\hline \noalign{\smallskip}                % inserts double horizontal lines
Parameter & This work & B19 \\    % table heading 
\noalign{\smallskip} \hline  \noalign{\smallskip}                      % inserts single horizontal line
           $x_1 \, ('')$ & $1.21^{+0.24}_{-0.24}$ & $1.40^{+0.23}_{-0.23}$   \\
            \rule{0pt}{2.5ex}
            $y_1 \, ('')$ & $-0.69^{+0.19}_{-0.19}$ & $-0.74^{+0.16}_{-0.17}$ \\
            \rule{0pt}{2.5ex}
            $e$ & $-0.63^{+0.01}_{-0.01}$ & $-0.63^{+0.01}_{-0.01}$     \\
            \rule{0pt}{2.5ex}
            $\theta_e \, (^{\circ})$ & $-38.79^{+0.25}_{-0.26}$ & $-38.75^{+0.22}_{-0.23}$    \\
            \rule{0pt}{2.5ex}
            $r_c \, ('')$ & $17.37^{+0.35}_{-0.37}$ & $18.06^{+0.53}_{-0.52}$  \\
            \rule{0pt}{2.5ex}
            $\sigma_{\mathrm{LT},1} \, (\mathrm{km \, s^{-1}})$ & $1159.1^{+6.8}_{-7.4}$ & $1162.4^{+6.4}_{-6.7}$  \\
            \rule{0pt}{2.5ex}
            $x_2 \, ('')$ & $-50.4^{+4.5}_{-5.6}$ & $-50.2^{+3.7}_{-4.4}$    \\
            \rule{0pt}{2.5ex}
            $y_2 \, ('')$ & $27.5^{+3.2}_{-2.7}$ & $26.8^{+2.4}_{-2.2}$ \\
            \rule{0pt}{2.5ex}
             $\sigma_{\mathrm{LT},2} \, (\mathrm{km \, s^{-1}})$ & $214^{+29}_{-26}$ & $221^{+24}_{-22}$  \\
 \noalign{\smallskip} \hline                                   %inserts single line
\end{tabular}
\end{table}
    
    In Table \ref{bestfit}, the best-fit values of the free parameters describing the cluster-scale DM haloes for this model and for that by B19 are listed. In the table, the index $1$ refers to the main, central halo, the index $2$ to the secondary, spherical one. The positions of the clumps are indicated with respect to the centre of the BCG. The ellipticity of the main halo is defined as $e= \frac{a^2-b^2}{a^2+b^2}$, where $a$ and $b$ are the major and minor semi-axes. The orientation angle is measured counter-clockwise from the positive $x$ axis. For both models, we determine the statistical uncertainties on the parameters using the Monte Carlo Markov chains as implemented by \texttt{LensTool}, based on the \texttt{BayeSys} algorithm \citep[][]{jullo07}. To obtain a reliable estimate of the uncertainties on the parameter values from the Monte Carlo sampling, the observational uncertainty on the positions of the multiple images has been increased to get a $\chi^2$ value close to the number of degrees of freedom.
    %This methods begins with a burn-in phase, a preliminary exploration of the parameter space, in which the posterior probability distribution is proportional to $(\mathcal{L})^{\lambda} P$, where $\mathcal{L}$ is the likelihood and $P$ is the prior on the parameter values. $\lambda$ is a cooling factor, which starts from zero and grows up to one with fixed steps, making the chains less sensitive to local maxima.
    
\section{Comparison and discussion}\label{s4}
   
   In this section we compare our model with that presented in B19, which is our starting point and reference. Using the same code, \texttt{LensTool}, B19 obtain a minimum value of $\chi^2$ of 69.90, which corresponds to a value of the rms on the positions of the 55 multiple images of $0.55''$, slightly lower than our value of $0.60''$. The maximum values of the likelihood for the two models, provided by \texttt{LensTool}, are reported in Table \ref{BICtable}. This small increase in the rms value with respect to B19 was expected, as in our model the mass component representing the member galaxies has less freedom. In B19, the truncation radius and the fiducial velocity dispersion of the $i$-th member are obtained through power laws from its luminosity $L_i$, specifically
   \begin{equation}\begin{split} \label{scaling}
       \sigma_{\mathrm{LT},i}& = \sigma^\mathrm{ref}_\mathrm{LT} \left(\frac{L_i}{L_0}\right)^{0.27}, \\
        r_{t,i}= & r^\mathrm{ref}_t \left(\frac{L_i}{L_0}\right)^{0.66},
   \end{split} \end{equation}
   where $L_0$ is the luminosity of the BCG. The reference values in the relations ($\sigma^\mathrm{ref}_\mathrm{LT}$ and $r^\mathrm{ref}_t$) are free to vary, increasing the number of free parameters from $49$ to $51$. This additional freedom corresponds to some extent to a possible mass exchange between the member galaxies and the diffuse cluster-scale DM haloes. In our model, the total mass component of the cluster galaxies is completely fixed by the observations through the FP, and only the two main DM haloes have their mass free to vary.

      \begin{table}
\caption{Maximum likelihood ($\mathcal{L}$),  and rms values for our model and for the model presented in B19.}             % title of Table
\label{BICtable}      % is used to refer this table in the text
\centering                          % used for centering table
\begin{tabular}{c c c}        % centered columns (4 columns)
\hline \noalign{\smallskip}                % inserts double horizontal lines
Parameter & This work & B19 \\    % table heading 
\noalign{\smallskip} \hline  \noalign{\smallskip}                      % inserts single horizontal line
            Observables & $110$ & $110$   \\
            Free parameters & $49$ & $51$ \\
            $\ln \mathcal{L}$ & $-65.94$ & $-58.43$    \\
            rms & $0.60''$ & $0.55''$  \\
 \noalign{\smallskip} \hline                                   %inserts single line
\end{tabular}
\end{table}
   
   \begin{figure}
  \centering
    \includegraphics[width=9cm]{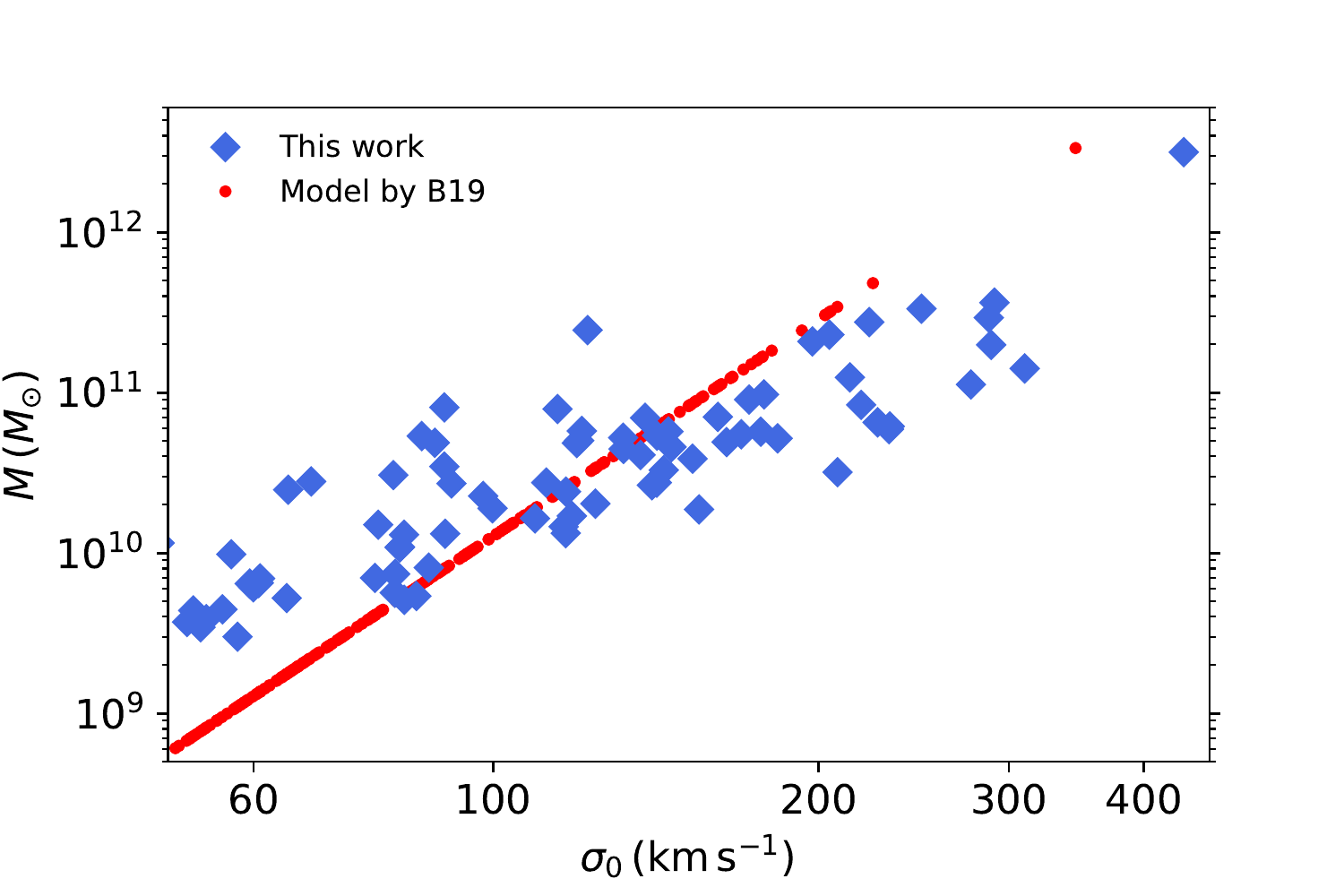}
  \caption{Total mass of the cluster members as a function of their velocity dispersion as predicted by our model (in blue) compared to values predicted by the model presented in B19 (in red).}
  \label{sigmamass}
  \end{figure}
   
   We first focus on how mass is assigned to the member galaxies. For both models, cluster members are modelled as dPIEs with vanishing core radius and zero ellipticity, so their total mass can be obtained as
   \begin{equation} \label{massPIEMD}
    M=\frac{\pi \sigma_0^2 r_t}{G}
   .\end{equation}
    In B19, $r_t$ and $\sigma_0$ are derived, for each member, with the two power-law scaling relations reported in Eq. \ref{scaling}, which imply that $M \propto \sigma_0^{4.44}$. Such a precise and simple relation between $M$ and $\sigma_0$, with no scatter, might be an oversimplification given the large range in total mass value and the great variety in galactic morphology. The use of two measured observables and a more accurate relation (the FP) should allow for a more complex and realistic dependence of $M$ from the other observables. This is shown in Fig. \ref{sigmamass}: the $M$($\sigma_0$) relation has a visible scatter, which the model by B19 could not include. The relation also shows a shallower slope (a bi-logarithmic fit on the members with $\sigma_{0,\mathrm{FP}} > 80 \, \mathrm{km \, s^{-1}}$ gives a slope value of around $2.4$). The two models are similar in the high-mass regime and agree on the prediction of the total mass of the BCG. Despite being fixed by the FP with the same procedure as for all other cluster members, our model predicts a BCG total mass value of $3.16 \times 10^{12} \, M_{\odot}$, very close to that found in B19 ($3.35 \times 10^{12} \, M_{\odot}$), where it is de facto a free parameter. Furthermore, the new procedure also leads to a different degeneracy between the parameters that describe the total mass of the sub-haloes. While, as anticipated in Sect. \ref{s3}, the introduction of a kinematic prior in B19 already allows for a significant reduction in the degeneracy between the values of $r_t$ and $\sigma_0$ for the various cluster members, the two reference values that appear in Eq. \ref{scaling} are still clearly degenerate (as shown in Fig. 4 from B19). In this work, instead, the values of $\sigma_0$ are fixed a priori from observations, while the values of $r_t$ are obtained at a later stage, testing several values for the proportionality constant with respect to the observed values of the effective radius.
    
    \begin{figure}
   \centering
   \includegraphics[width=9cm]{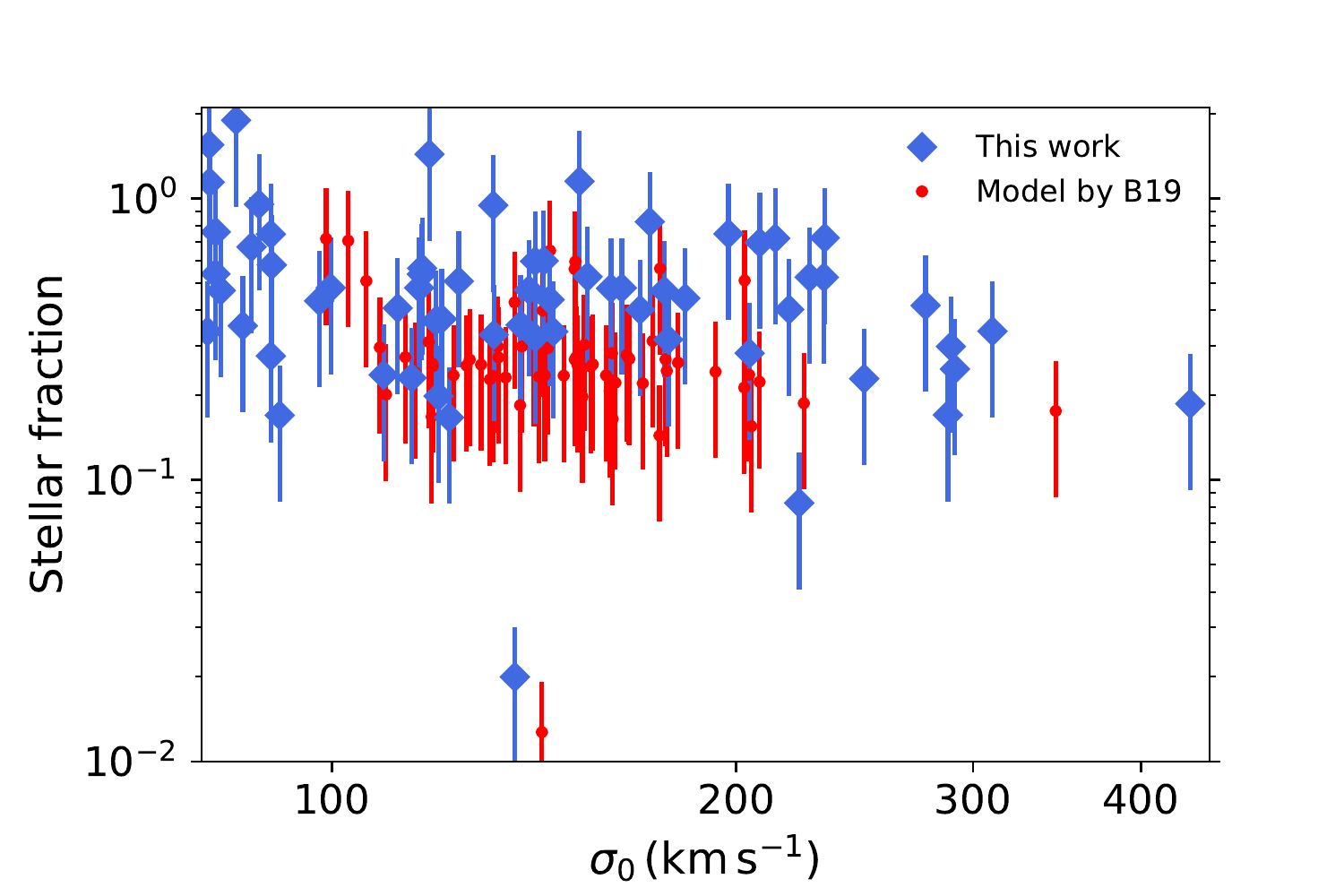}
          \caption{Stellar fraction of the cluster members as a function of their central velocity dispersion for the 63 galaxies with $\sigma_{0,\mathrm{FP}} > 80 \, \mathrm{km \, s^{-1}}$: comparison between the values obtained with our model (in blue) and those from the model in B19 (in red).}
         \label{fraction}
   \end{figure}
   
   As explained in Sect. \ref{s2}, we measured the stellar mass values of all 222 member galaxies \citep[][]{mercurio21} from HST multi-band photometric data, using a Salpeter IMF. By combining this with the results of Eq. \ref{massPIEMD}, we can estimate the stellar fraction of the cluster members, computed as the ratio between their stellar mass and total mass values. As lensing models are more sensitive to the effects of high-mass members, we focused only on the 63 galaxies with $\sigma_{0,\mathrm{FP}} > 80 \, \mathrm{km \, s^{-1}}$.
   For this subset of members, the median value of the stellar fraction is $0.47$, with a standard deviation of $0.34$. From the model presented in B19, using the same stellar mass values, we obtain a value of $0.25$. Therefore, our model predicts, on average, slightly higher values for the stellar fraction, as illustrated in Fig. \ref{fraction}.
   
\begin{figure}
   \centering
   \includegraphics[width=9cm]{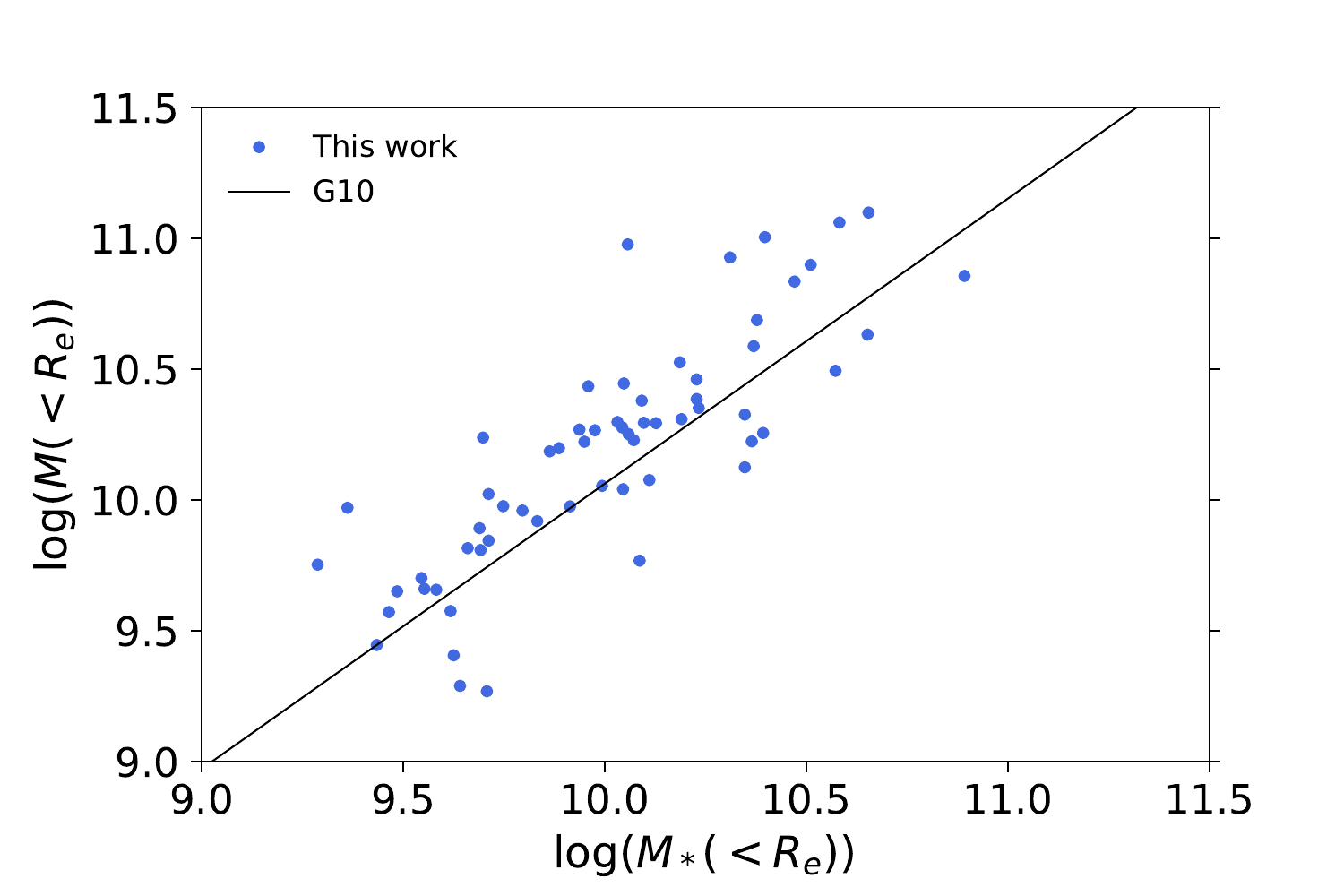}
      \caption{Logarithmic values of the total mass as a function of the logarithmic values of the stellar mass, both projected within the effective radius, of all members with $\sigma_{0,\mathrm{FP}} > 80 \, \mathrm{km \, s^{-1}}$. The plot also shows the relation found by \citet{grillo10}.}
         \label{grfig}
   \end{figure}

   We now compare our stellar fraction values with those of other massive early-type galaxies. \citet{grillo10} studied the projected DM fraction within the effective radius for a sample of nearly $1.7\times 10^5$ early-type galaxies from the Sloan Digital Sky Survey (SDSS) Data Release Seven (DR7), selected to reproduce well the physical properties of the lenses in the SLACS survey \citep{koopmans09}. He considered the values of the galaxy stellar mass as reported in the Max Planck for Astrophysics and John Hopkins University catalogue, and obtained the values of the galaxy projected total mass within the effective radius, by assuming a one-component isothermal model, parametrised by the value of the measured central stellar velocity dispersion. He found that the best-fit linear relation between the logarithm of the projected total and stellar mass values within the effective radius, $M(<R_e)$ and  $M_*(<R_e)$, respectively, is
   \begin{equation}\label{grillo10}
       \log (M(<R_e)) = -0.58 + 1.09 \log (M_*(<R_e)).\end{equation}
   As far as the cluster members of AS1063 are concerned, assuming that light traces stellar mass, $M_*(<R_e)$ is half of their total stellar mass, as $R_e$ is defined as the two-dimensional radius within which half of the total light is emitted. The stellar mass values of our cluster members have been obtained assuming a Salpeter stellar IMF, while those presented by \citet{grillo10} were derived adopting a Chabrier stellar IMF \citep[][]{chabrier03}. To compare the two samples, we use a constant conversion factor \citep[$0.585$, from][]{speagle14} between the stellar mass values obtained with a Salpeter and a Chabrier stellar IMF. As for the projected total mass, the mass of a dPIE profile with a vanishing core radius enclosed within a two-dimensional radius $R$ is \citep[][]{eliasdottir07}
   \begin{equation}\label{projmass}
       M(<R) = \frac{\pi \sigma_0^2}{G} \left(R-\sqrt{r_t^2+R^2}+r_t\right).
   \end{equation}
   For $r_t=2.3 R_e$, Eqs. \ref{projmass} and \ref{massPIEMD} give $M(R_e) \approx 0.344 M$, where $M$ is the total mass. As Fig. \ref{grfig} shows, we find that our members follow well the relation found by \citet{grillo10}. This is significant, as we are considering in both cases massive elliptical galaxies, but the values of their stellar fractions are obtained with different methods. The observed agreement suggests that the inner (i.e. within $R_e$) mass structure of the massive cluster members of AS1063 and of massive early-type galaxies living in different environments might be very similar. It should be noted that the relation shown in Fig. \ref{grfig} has been adapted from Eq. \ref{projmass} for the case of stellar mass values determined using a Salpeter IMF. 
   
   \begin{figure}
   \centering
   \includegraphics[width=9cm]{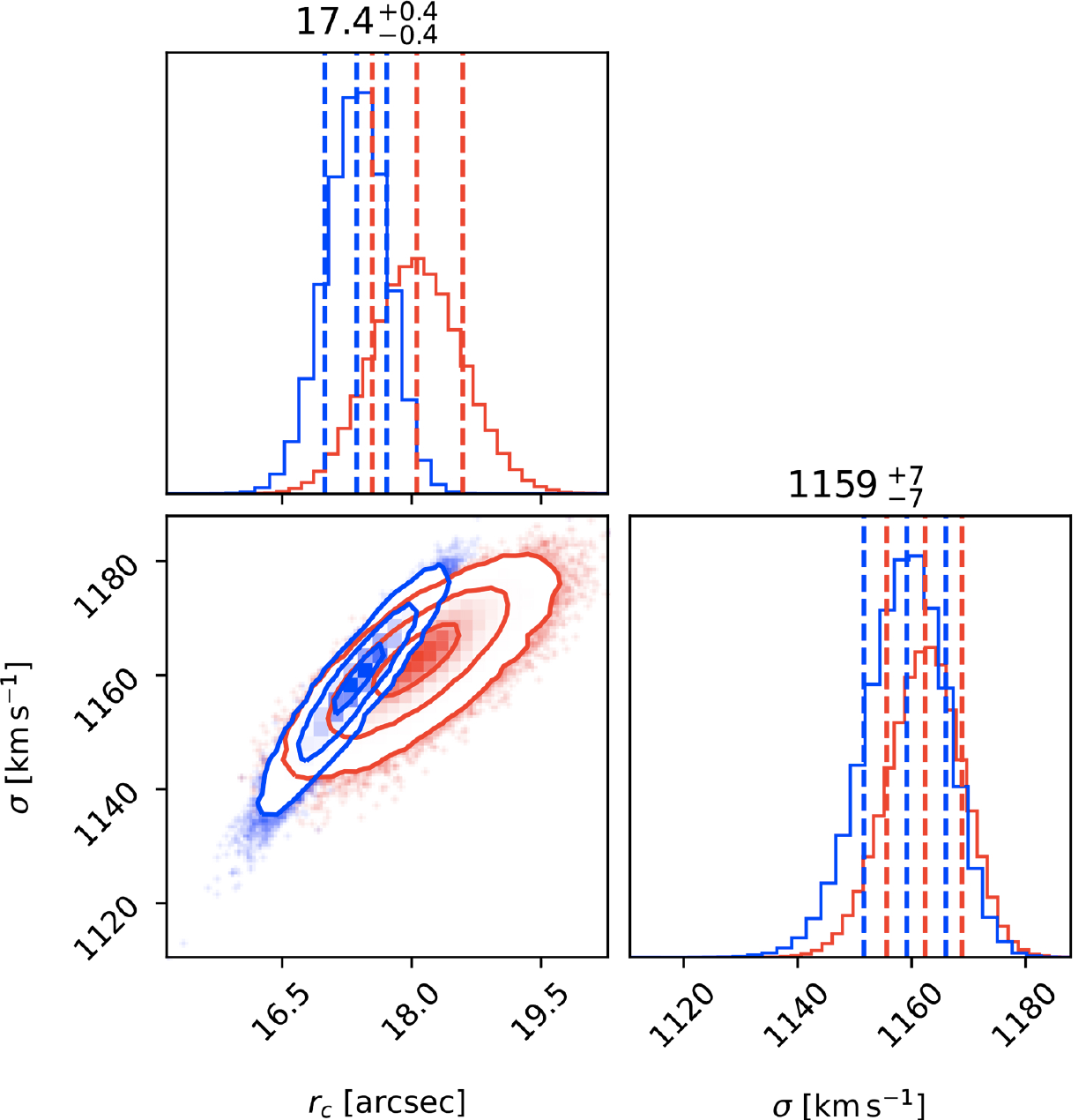}
      \caption{Posterior probability distributions of the core radius, $r_c$, and fiducial velocity dispersion, $\sigma_{\mathrm{LT}}$, values of the main DM halo and degeneracy between them for our model (in blue) and that from B19 (in red). The $16^\mathrm{th}$, $50^\mathrm{th}$, and $84^\mathrm{th}$ percentiles of the marginalised probability distributions are marked with dashed vertical lines for both models. The values of these percentiles for our model are reported above the histograms.}
         \label{degcompzoom}
   \end{figure}
   
   Looking instead at the diffuse cluster-scale components of the two models, from the Monte Carlo sampling of the marginalised posterior probability we see a drop of more than $30\%$ of the uncertainty on the value of the core radius of the main DM halo (see Fig. \ref{degcompzoom}). The scatter between the core radius and the velocity dispersion is also considerably reduced. This indicates that a more accurate estimate of the total mass of the member galaxies has resulted in distributions of the parameters with a lower statistical uncertainty.
   %partly limiting the possible mass exchange between the different mass components
   
   Monte Carlo Markov chains also show that the core of the main diffuse DM halo has a radius of $86 \pm 2 \, \mathrm{kpc}$. Our result is in line with those from SL models of similar galaxy clusters \citep[e.g. B19,][]{grillo15,annunziatella17,lagattuta17,caminha19,diego20}, which found that cluster haloes are very well described by cored isothermal mass density profiles. Figure \ref{degcompzoom} also shows a clear degeneracy, with a positive correlation, between the velocity dispersion and the core radius of the main DM halo. This is expected, as different combinations of these two parameters can determine the same value of the projected mass enclosed within a certain radius from the cluster centre. In particular, a higher value of central density scale, here included in the definition of $\sigma_\mathrm{LT}$, can be compensated by a larger core in the density profile, and vice versa, to obtain the same mass within the multiple images of the same family.
   
   Finally, from Table \ref{bestfit}, we notice that the values of our and B19 best-fit parameters of the diffuse DM component are very similar. The only significant discrepancy is the value of the core radius of the main DM halo of the cluster. This can be explained by considering that, far from the cluster centre, the total mass distribution of the cluster is dominated by the cluster-scale DM halo component. The positions of the multiple images observed in these regions provide strong constraints on the values of all the parameters of this halo. The core radius of the cluster DM halo is more sensitive to the total mass profile of the cluster close to its centre, which is more directly related to the mass distribution of the cluster members located in this region. A change in the mass contribution of the cluster members is therefore reflected in a small change of the DM halo core radius.
  
\section{Mass profile decomposition}\label{s5}

   \begin{figure*}
   \centering
   \includegraphics[width=19cm]{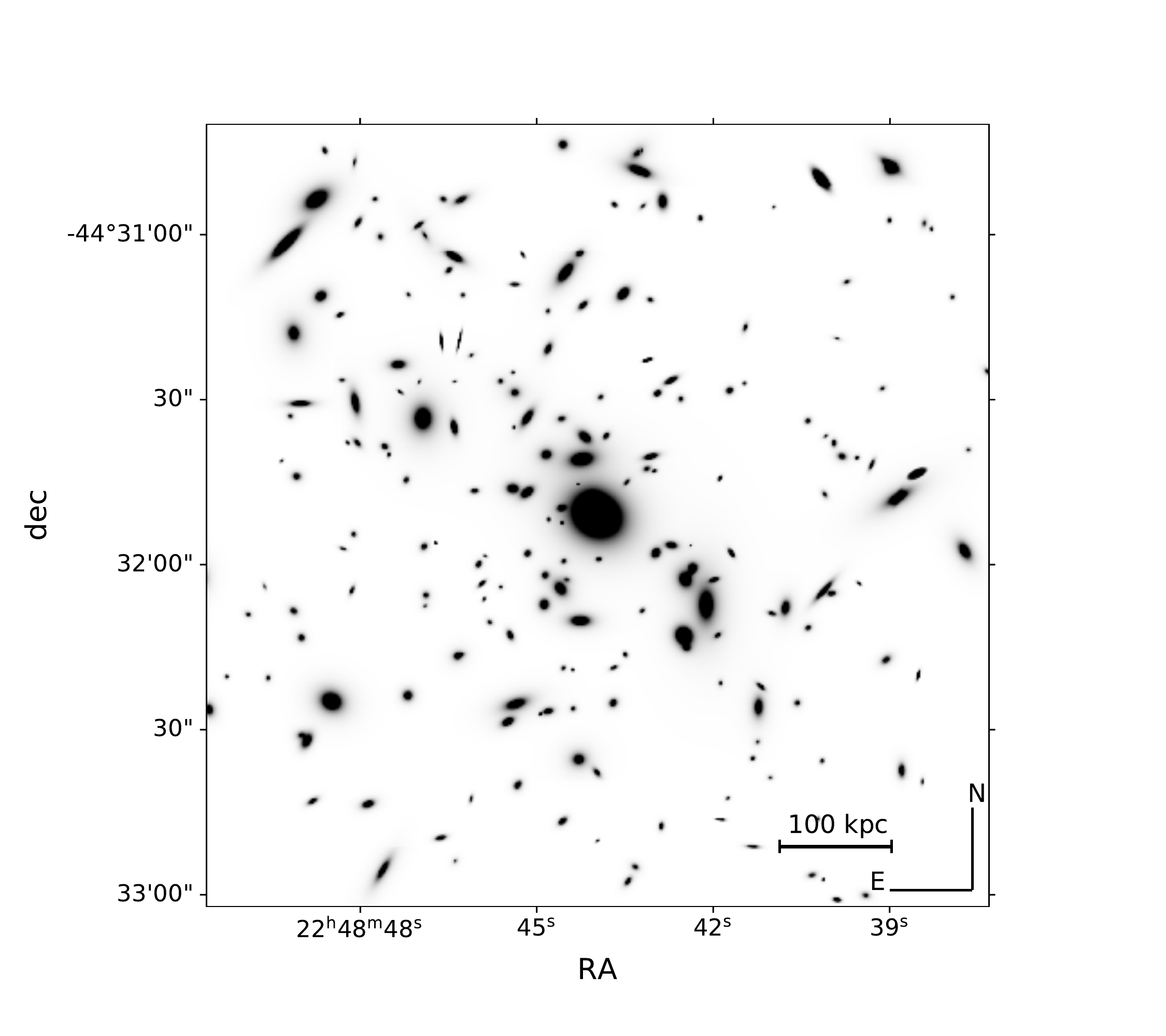}
      \caption{Reconstructed surface brightness distribution of all cluster members used in the SL model in the HST F814W band. The field of view is 700 kpc wide and centred on the BCG to cover the entire region where the cumulative mass profile is obtained.}
         \label{f814hst}
   \end{figure*}

   \begin{figure*}
   \centering
 %  \hspace{-10mm}
%       \begin{minipage}[c]{.40\textwidth}
\includegraphics[scale=0.58] {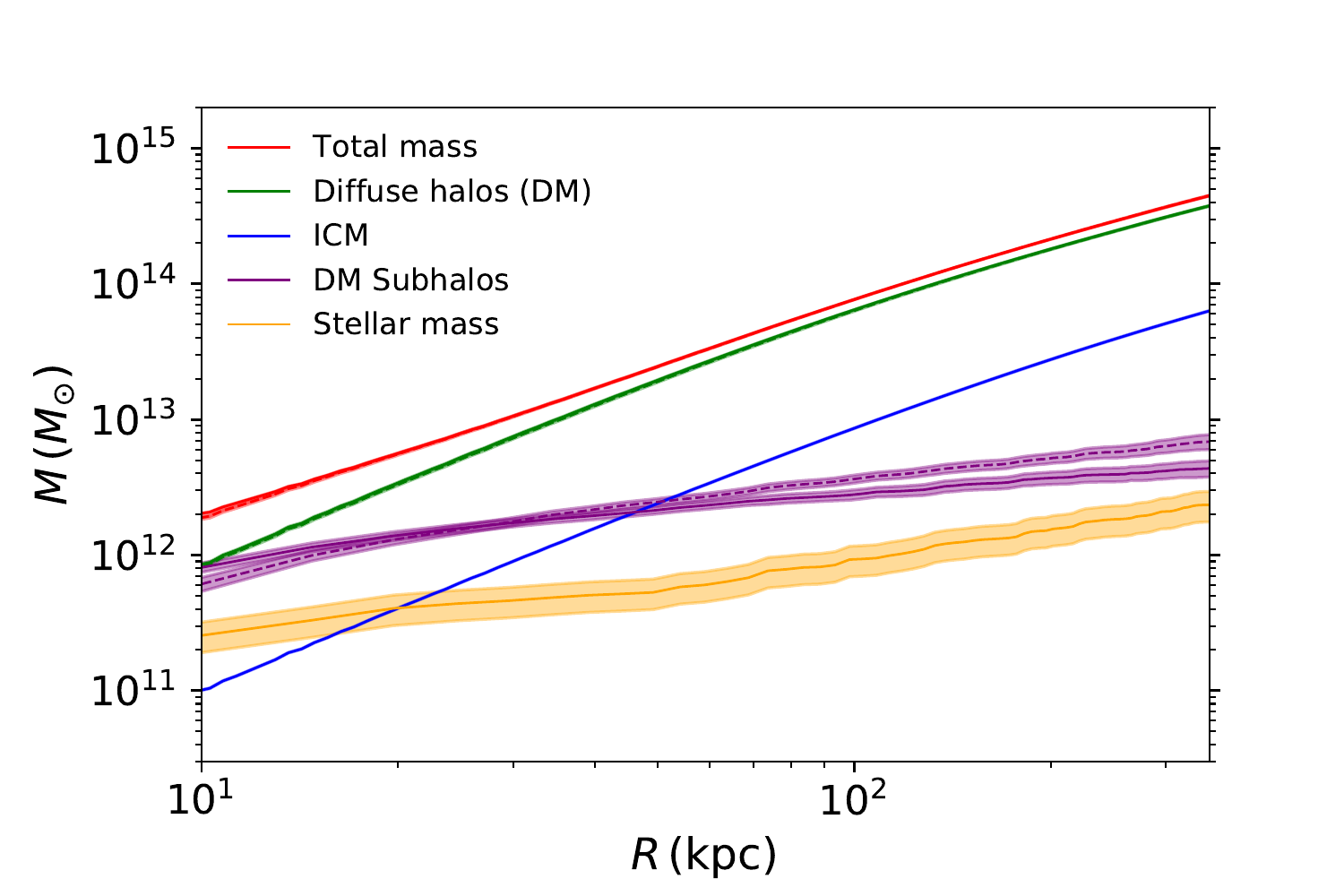} 
%       \end{minipage}
 %  \hspace{15mm}
%       \begin{minipage}[c]{.40\textwidth}
\includegraphics[scale=0.58] {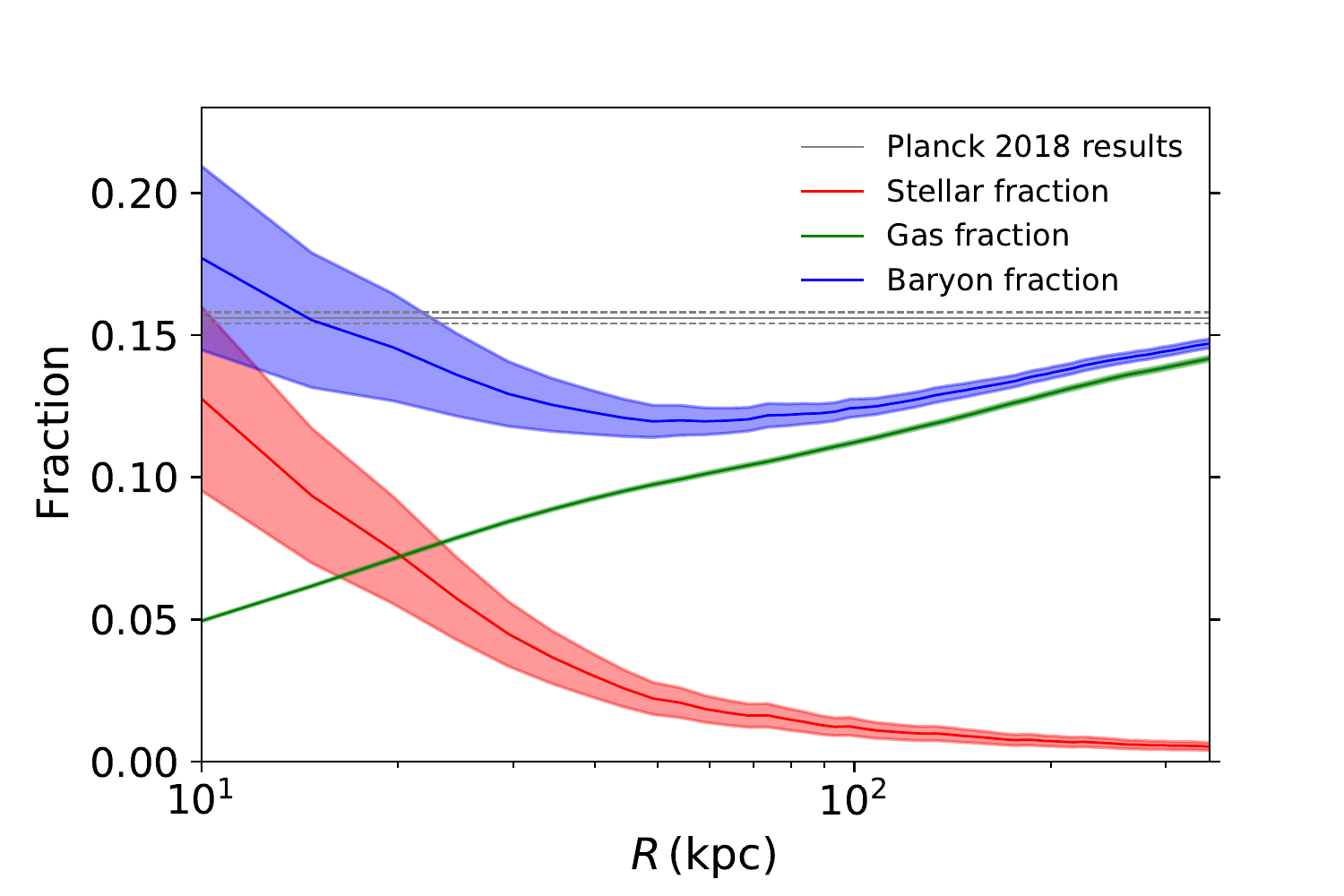} 
        
%       \end{minipage}\label{bbb}
   \caption{Cumulative mass and mass fraction profiles of the cluster components out to a projected radius of $350 \, \mathrm{kpc}$. Left panel: Cumulative two-dimensional mass profiles for all baryonic and dark components of AS1063. The total mass profile and those of the diffuse DM haloes and DM sub-haloes obtained from the model in B19, which differ from ours, are represented with dashed lines. Right panel: Cumulative stellar-, gas-, and baryonic-to-total mass fractions from our model. The value of the cosmological baryon fraction from the \citet{planck20} is also shown.}
   \label{bbb}
   \end{figure*}
   
   In this section we obtain and present the cumulative projected mass profiles of all the cluster mass components: the stellar mass and the DM of cluster members, the ICM, and the diffuse, cluster-scale DM. We only consider the mass distribution out to a projected radius $R$ of $350 \, \mathrm{kpc}$ from the centre of the BCG, to avoid including regions not covered by our photometric data, where the cluster member catalogue is incomplete.
   
   \subsection{The baryonic mass components}
   
   To measure the stellar mass profile of the cluster, we started from the surface brightness best-fit model of each of the cluster members in the HST F814W band, shown in Fig. \ref{f814hst}, which we already used to derive their structural parameters. Figure \ref{f814hst} shows that our procedure does not allow the intra-cluster light distribution, which also contributes to the cluster stellar mass budget, to be completely accounted for. However, the intra-cluster light  in AS1063, especially far from its BCG, is less important than in other clusters of comparable mass. Similarly to \citet{annunziatella17}, we estimated an average value for the stellar mass-to-light ratio ($M_*/L$) of all cluster members within a projected radius of $350 \, \mathrm{kpc}$ from their total luminosity and total stellar mass. The median value of the stellar mass-to-light ratio for these members is $1.0 \, M_\odot/L_\odot$, estimated in the HST F814W band.
   The choice of considering a constant ratio is an approximation, as the value may change in the different regions of each galaxy, as well as among galaxies. However, several galaxies are included in each radial bin above $50 \, \mathrm{kpc}$, and this should reduce the impact of this approximation. We then used this ratio to convert the cumulative luminosity profile into a cumulative stellar mass profile. As above, we considered an error of $0.22$ on $\log(M_*)$ for each galaxy, and we applied the same uncertainty to the total value in each radial bin of the projected profile.
   
   We take the cumulative projected ICM mass profile from \citet{bonamigo18}: following their result, we impose a relative error of $1\%$ on each bin. The total projected mass of the three haloes used to model the hot gas, within a bi-dimensional radius of $350 \, \mathrm{kpc}$, is reported in Table \ref{massprof}.
   
         \begin{table}
\caption{Projected mass within a two-dimensional radius $R=350 \, \mathrm{kpc}$ of the baryonic components.}             % title of Table
\label{massprof}      % is used to refer this table in the text
\centering                          % used for centering table
\begin{tabular}{c c}        % centered columns (4 columns)
\hline \noalign{\smallskip}                % inserts double horizontal lines
Component & $M(10^{12} \, M_\odot)$ \\    % table heading 
\noalign{\smallskip} \hline  \noalign{\smallskip}                      % inserts single horizontal line
           Stars & $2.4 \pm 0.6$   \\
            \rule{0pt}{2.5ex}
            ICM & $63.3 \pm 0.6$ \\
\noalign{\smallskip} \hline 
\end{tabular}
\end{table}

    \begin{table}
   \caption{Comparison between the projected mass within a two-dimensional radius $R=350 \, \mathrm{kpc}$ for the DM components in our model and in the model from B19.}             % title of Table
   \label{massprofdark}      % is used to refer this table in the text
   \centering                          % used for centering table
   \begin{tabular}{c c c}        % centered columns (4 columns)
   \hline \noalign{\smallskip}                % inserts double    horizontal lines
    Component & Our model & Model from B19 \\    % table heading 
     & $M(10^{12} \, M_\odot)$ & $M(10^{12} \, M_\odot)$ \\
    \noalign{\smallskip} \hline  \noalign{\smallskip}                      % inserts single horizontal line
           Diffuse DM & $377 \pm 2$  &  $377 \pm 2$  \\
            \rule{0pt}{2.5ex}
            DM sub-haloes & $4.4 \pm 0.6$ & $6.9 \pm 0.9$  \\
             \rule{0pt}{2.5ex}
            Total & $447 \pm 2$ & $449 \pm 2$ \\
   \noalign{\smallskip} \hline                                     %inserts single line
   \end{tabular}
   \end{table}
   
   \subsection{The dark matter mass distribution}
   
   In Sect. \ref{s3}, we described our modelisation of the DM mass distribution: an ellipsoidal, pseudo-isothermal, non-truncated clump, and a less massive, spherical one represent the cluster-scale DM haloes. On the other hand, $222$ spherical, truncated, isothermal haloes were introduced to describe the total mass of cluster members. To disentangle the mass distribution of DM sub-haloes, we therefore need to subtract the stellar mass profile to it.
   
   The parameter space sampling obtained with the Monte Carlo method can be used to estimate the error on the DM cumulative mass profiles. We extracted 100 random sets of parameter values from the chains and derived the cumulative mass profiles of each corresponding model. In Fig. \ref{bbb} and in Table \ref{massprofdark}, the error bars show the $16^\mathrm{th}$ and $84^\mathrm{th}$ percentiles of the mass profiles associated with all the models extracted from the chains. We performed this analysis for our model and repeated it for the model by B19. In the first case, the total mass of the cluster members was fixed, so the uncertainty on the DM sub-halo mass profile is only associated with the error on the stellar mass. In the second case, instead, the mass of the cluster members was also optimised: Monte Carlo Markov chains provided us with an uncertainty on their mass, which was combined with that on $M_*$ to obtain the error on the DM sub-halo cumulative mass profile.
   
   We notice that the total mass of the cluster and the mass of the diffuse DM component are very well constrained by SL: the two models find very similar values and their statistical uncertainties are around $1\%$. As a matter of fact, the two profiles are almost completely superimposed in Fig. \ref{bbb}, and the error bars are very small. The mass of the DM sub-haloes, instead, slightly differs between the two models, especially in the outer regions, where our model predicts a lower value. A reason for this discrepancy might be in the different scaling laws that determine $r_t$ in the two models. Comparing our results with the total mass profiles obtained, for the same cluster, with a full dynamical reconstruction and with the X-ray hydrostatic method \citep[][]{sartoris20}, and with combined weak and SL \citep[][]{umetsu16}, we find very good agreement in the radial range considered. These two profiles are shown in Fig. 7 by \citet{sartoris20}. 
   
   The profiles measured from our model also allow us to obtain the two-dimensional, cumulative, stellar-, gas-, and baryonic-to-total mass fractions as a function of $R$, out to $350 \, \mathrm{kpc}$. In Fig. \ref{bbb}, we report the results and compare the baryonic fraction profile with the cosmological baryon fraction value reported in the Planck 2018 results: $0.156 \pm 0.002$, obtained from the ratio between $\Omega_b h^2$ and $\Omega_m h^2$, with information from the cosmic microwave background power spectrum and lensing reconstruction \citep[see Planck Collaboration][]{planck20}. The error bars of the profiles result from the propagation of the uncertainty on the cumulative values of the mass components. The fraction obtained at $R=350 \, \mathrm{kpc}$ is $0.147 \pm 0.002$. This value is similar to the cosmological value and consistent with what \citet{bonamigo18} reported for the same cluster at a comparable radius.
   
\section{Comparison with halo occupation distribution studies}\label{s6}

  \begin{figure*}
   \centering
 %  \hspace{-10mm}
%       \begin{minipage}[c]{.40\textwidth}
\includegraphics[scale=0.58] {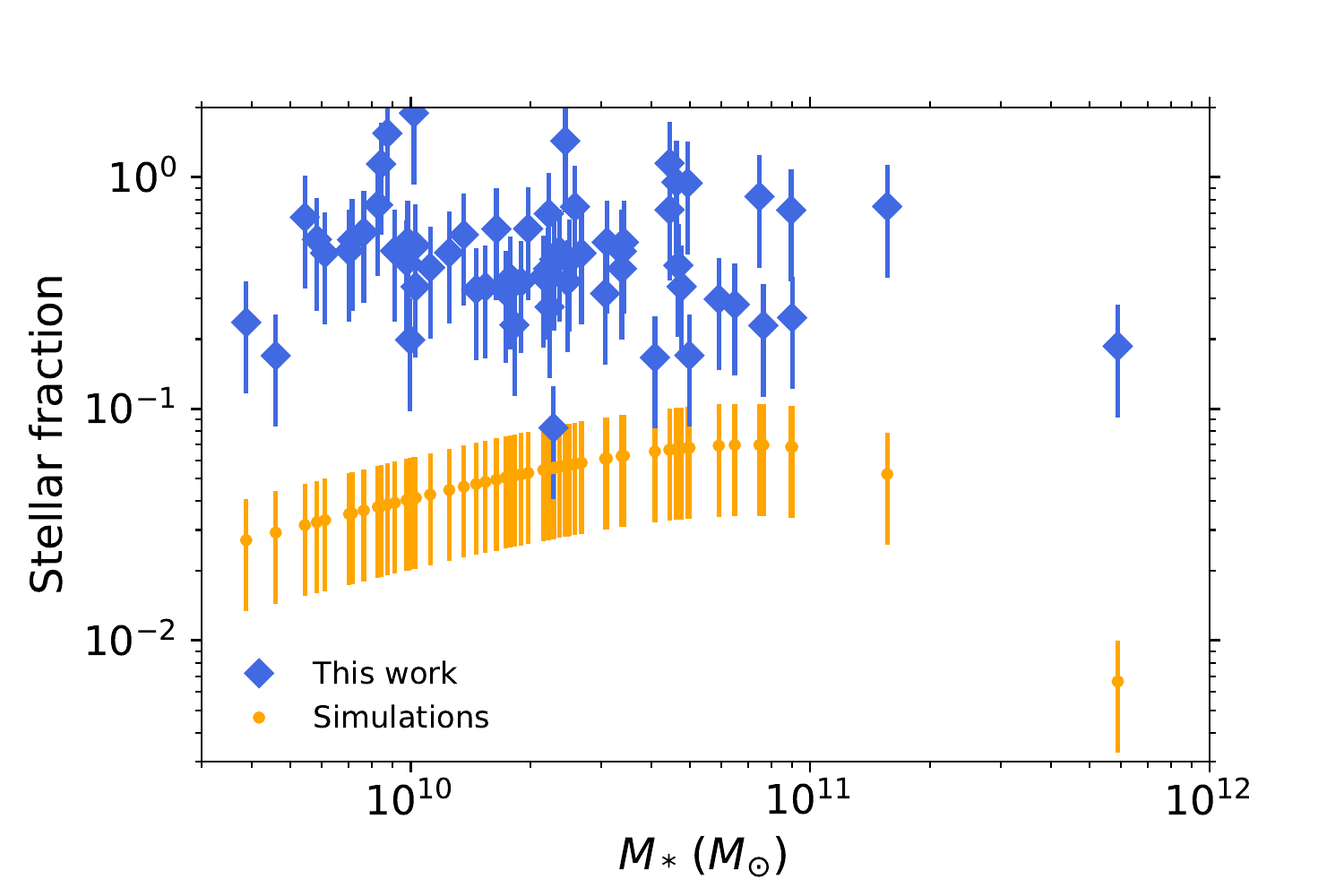} 
%       \end{minipage}
 %  \hspace{15mm}
%       \begin{minipage}[c]{.40\textwidth}
\includegraphics[scale=0.58] {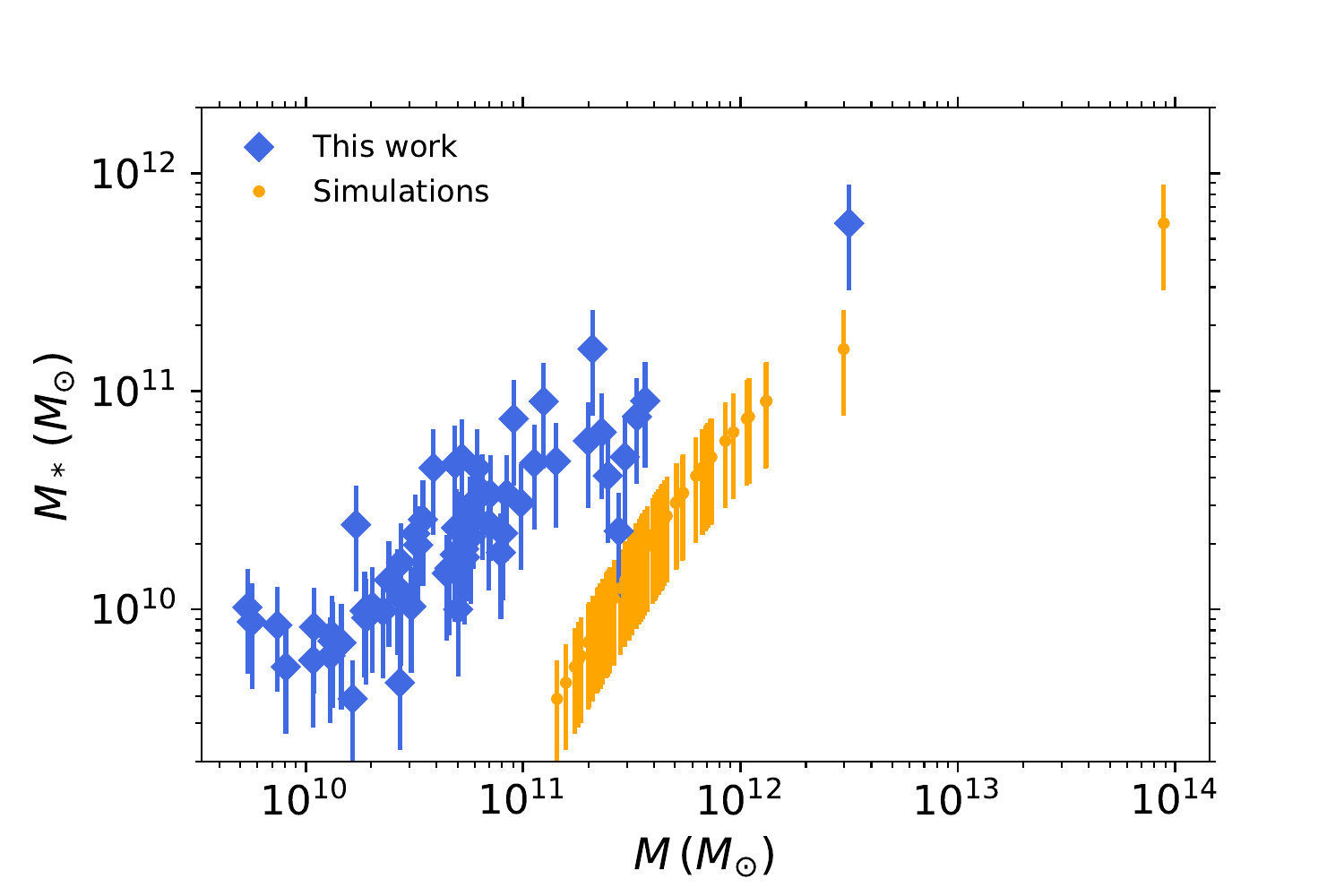} 
        
%       \end{minipage}\label{bbb}
   \caption{Comparison between the values of the stellar over total mass fraction of the cluster members obtained with our lensing model (in blue) and with the SHMR from \citet{girelli20} (in orange). Only the 63 cluster members with $\sigma_{0,\mathrm{FP}} > 80 \, \mathrm{km \, s^{-1}}$ are shown. Left panel: Stellar over total mass fraction of members as a function of their stellar mass. Right panel: SHMR.}
   \label{aaa}
   \end{figure*}

   In this section we compare the values of the total mass predicted for cluster members by our model with those obtained from the measured stellar mass values using halo occupation distribution (HOD) studies, based on cosmological $N$-body simulations. 
   In the previous section, we found that both our model and that from B19 lead to median values of the stellar-to-total mass ratio of massive member galaxies of $0.25$ at least. As we will see, this result differs significantly from the predictions of statistical (sub)-halo abundance matching studies. We compare our results with the recent stellar-to-halo mass relation (SHMR) presented in \citet{girelli20}. They considered the halo mass function derived from the cosmological $N$-body collisionless $\mathrm{\Lambda}$CDM DUSTGRAIN-\textit{pathfinder} simulation \citep[][]{giocoli18}, which is complete for halo masses $M_h>10^{12.5} \, M_\odot$, and extended it to lower masses. The simulation is built from a sample that also includes some cluster members, especially in the high-mass range. Then, taking the observed galaxy stellar mass of the COSMOS field \citep[][]{scoville07}, they performed an abundance matching via a deterministic relation between halo and stellar mass. The SHMR is found by \citet{girelli20} to be in good agreement both with the observed clustering of massive galaxies and with the SHMR of clusters selected from spectroscopic data from the SDSS. The inclusion of some scatter on the value of the stellar mass leads to consistent results for the relation, so we do not consider it in the following analysis. 
   
   The obtained SHMR is valid from stellar mass values between approximately $10^8 \, M_\odot$ and $10^{12} \, M_\odot$, from $z=0$ to $z=4$. The relation, using the parametrisation proposed by \citet{moster10}, can be written as
   \begin{equation} \label{SHMR}
    \frac{M_*}{M_h}(z)=2A(z)\left[\left(\frac{M_h}{M_A(z)}\right)^{-\beta(z)}+\left(\frac{M_h}{M_A(z)}\right)^{\gamma(z)}\right]^{-1},
   \end{equation}
   where $A(z)$ is a normalisation factor, $M_A(z)$ a characteristic halo mass, and $\beta(z)$ and $\gamma(z)$ are the slopes of the relation in the low- and high-mass ranges, respectively. We adopted the values of the parameters found by \citet{girelli20} at the cluster redshift. We then proceeded by using Eq. \ref{SHMR} to compute the total halo mass of the members and, in turn, the values of their truncation radii from Eq. \ref{massPIEMD}. We finally compared these values of the truncation radii with those we obtained in our model from the effective radii of the members. We did not consider any uncertainty on the parameter and stellar mass values, as we will show that the differences are large and cannot be reconciled by the statistical errors. We notice that the SHMR presented by \citet{girelli20} was derived adopting a Chabrier stellar IMF. To make use of the relation to estimate the halo mass of the cluster members from their stellar mass, we adopted the same conversion factor of 0.585 between the Salpeter and Chabrier IMFs that we previously used.
   
   Using the SHMR, the projected mass of the DM sub-haloes within a bidimensional radius of $350 \, \mathrm{kpc}$ grows to $6.1^{+0.1}_{-0.1} \times 10^{13} \, M_{\odot}$, larger by a factor of almost $14$. The ratio $r=\frac{M_{\mathrm{SHMR}}}{M_{\mathrm{model}}}$, where $M_{\mathrm{SHMR}}$ is the total mass obtained for a member from SHMR, and $M_{\mathrm{model}}$ is the corresponding total mass as in the SL model, has a median value of $8.6$ for the members with $\sigma_{0,\mathrm{FP}}>80$ $\mathrm{km \, s^{-1}}$. For these selected high-mass members, as shown in Fig. \ref{aaa}, the median value of the stellar fraction decreases from $0.47$ to $0.055$. The value of $r$ is also that of the ratio between the value of the truncation radius of a sub-halo obtained from simulations and that adopted in our model.
   
   We performed a new run of the model, fixing the total mass of each cluster member to the value obtained from the SHMR. Optimising the free parameters of the two diffuse DM haloes, the value of the rms between the observed and predicted positions of the $55$ multiple images grows by more than $25 \%$ (to $0.76''$), despite having the same number of degrees of freedom.  
  
    These results seem to indicate a significant discrepancy between SL models and the predictions of an SHMR derived from $N$-body cosmological simulations. The total halo mass values predicted by the SHMR are significantly higher than what is predicted by both our SL model and that from B19, and their inclusion in an SL model leads to a significant decrease in its accuracy.
    An important reason for this discrepancy may be the fact that the $\mathrm{\Lambda}$CDM DUSTGRAIN-\textit{pathfinder} simulation only includes collisionless particles. Thus, it does not consider the effects of the interplay between DM and baryons, which are very important during galaxy formation. Several complex phenomena taking place on small spatial scales, such as star formation, radiative gas cooling, and active galactic nuclei feedback can have a non-negligible effect on the physical properties of galactic haloes \citep[see][]{kravtsov12, planelles14,barnes17}.
    Moreover, the SHMR presented by \citet{girelli20} describes both cluster and field galaxies with the same law. During the formation of a cluster, galaxies fall into the cluster potential well. At first, they undergo a phase of star formation triggered by shocks in their gas component. At a later stage, both the baryonic and DM components of the members are stripped by ram pressure and gravitational tidal interactions \citep[see, for instance,][]{mccarthy08,vandenbosch08}. This phase could determine an SHMR different from that of field galaxies: \citet{smith16}, for instance, found that the DM stripping happens earlier than star stripping, leading to an increase in the ratio between the effective and virial radii of the sub-haloes. This could justify a higher value of the stellar mass fraction of cluster members when compared to field galaxies.
    While the most massive galaxies considered by \citet{girelli20} to calibrate the SHMR are mostly hosted in clusters and groups, the low-mass galaxies considered are mainly field galaxies, which have evolved in an environment different from that of the cluster members we focus on here.
    To address both the role of baryons and DM interactions and the influence of the dense environment of clusters in the evolution of galaxies, \citet{sirks21} used the Hydrangea/C-EAGLE suite \citep[][]{bahe17,barnes17} cosmological hydrodynamical simulations to build two distinct SHMRs, introducing a distinction between cluster and field galaxies. As clear from Fig. 5 in \citet{sirks21}, the SHMR of cluster members yields stellar-to-total mass fraction values that are almost an order of magnitude higher than those of field galaxies, at a fixed stellar mass. As we see in the next section, such an effect will alleviate the reported discrepancies between the stellar fraction values predicted by SHMRs and reconstructed from our SL model.
   
\section{Comparison with cosmological hydrodynamical simulations}\label{s7}
   
   In this section we contrast the features of the sub-haloes in our new SL model with the predictions of the most recent cosmological hydrodynamical simulations. In the first subsection we consider the stellar over total mass fraction of sub-haloes. In the second subsection we study the relation between their maximum circular velocity and total mass, which is an indication of their compactness \citep[][]{munari16}.
   
   \subsection{Stellar fraction of cluster members}
   
\begin{figure}
  \centering
    \includegraphics[width=9cm]{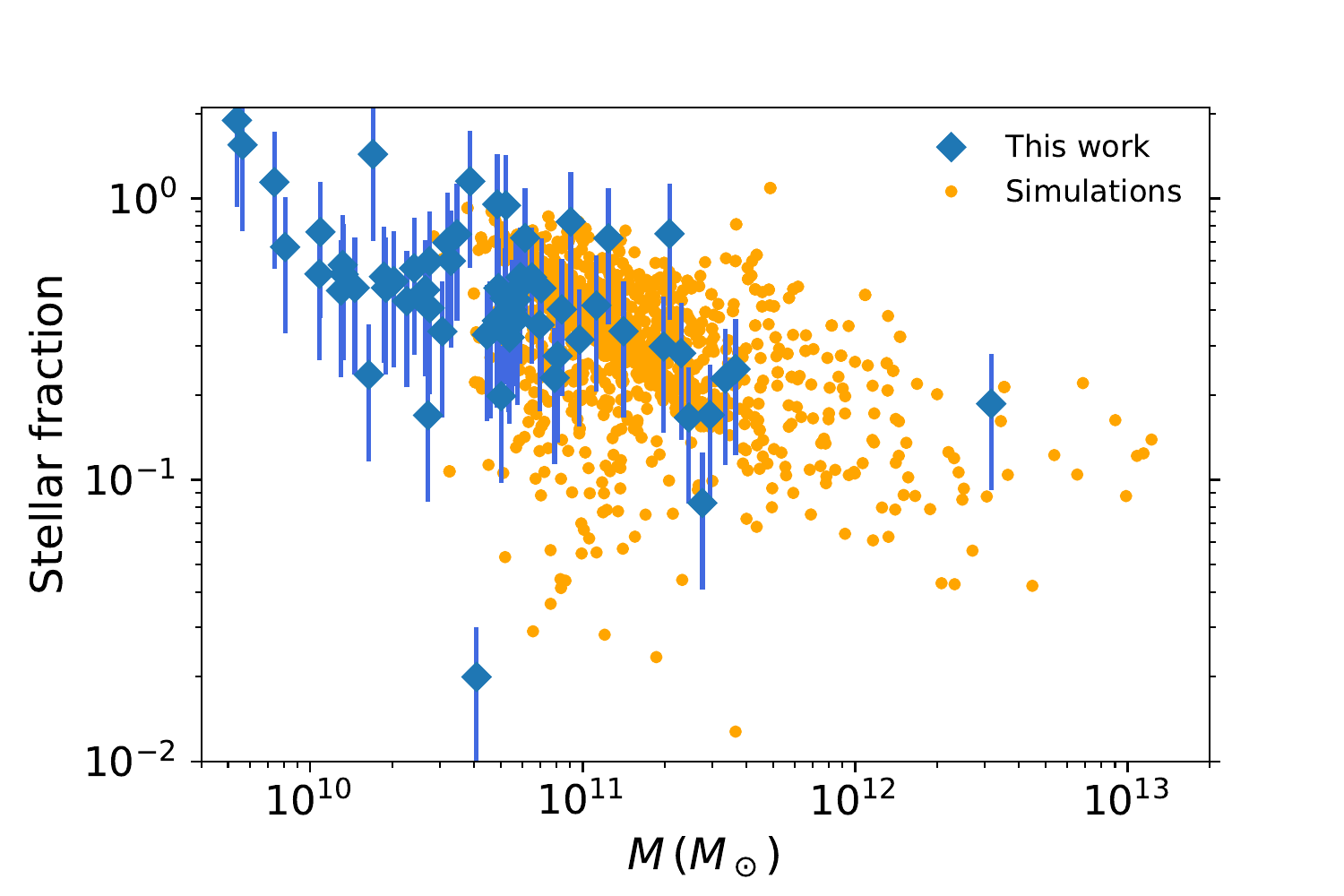}
  \caption{Comparison between the values of the stellar over total mass fraction of sub-haloes with velocity dispersion greater than $80 \, \mathrm{km \, s^{-1}}$ as a function of their total mass. The values reconstructed from our model are shown in blue and the values predicted via hydrodynamical simulations by \citet{planelles14} in orange. }
  \label{fractionsim}
  \end{figure}

   As reported in Sect. \ref{s4}, our model predicts a median stellar-to-total mass fraction value of $0.47$, with a standard deviation of $0.34$, for the 63 cluster members with $\sigma_{0,\mathrm{FP}}>80$ $\mathrm{km \, s^{-1}}$. We have seen, in Sect. \ref{s6}, that an SHMR obtained with an HOD procedure on the results of $N$-body cosmological simulations underestimates this value with respect to our model by almost an order of magnitude. We then repeated the comparison with high-resolution cosmological hydrodynamical simulations \citep[][]{planelles14,rasia15} carried out with the GADGET-3 code \citep[][]{springel05}. To perform a comparison between the features of sub-haloes as predicted by the simulations and by our model, we considered $18$ simulated clusters with $M>10^{15} \, M_\odot$ at redshift $z=0.38$ and performed three spatial projections of their mass distributions. We then extracted their sub-haloes from the \texttt{SUBFIND} algorithm \citep[][]{springel01,doulag09} catalogue\footnote{The simulation snapshots and simulated sub-halo catalogues used are available at https://www.ict.inaf.it/index.php/DOI/122-ds-2020-1.}. We imposed a minimum value of the velocity dispersion of $80 \, \mathrm{km \, s^{-1}}$, as in the previous analyses carried out in this paper, and a maximum projected distance from the cluster centre of $R=0.15 r_\mathrm{vir}$, where $r_\mathrm{vir}$ is the cluster virial radius. We thus obtained a catalogue of $1756$ sub-haloes, and we computed the stellar mass fraction for each of them. Again, as the simulations considered use a Chabrier IMF to obtain the stellar mass of sub-haloes, we used the previously presented conversion factor to compare them with our measured stellar mass values, for which \citet{mercurio21} used a Salpeter IMF.
   We obtain a median stellar mass fraction value of $0.37$, with a standard deviation of $0.18$, in good agreement with the value suggested by the lensing model. Figure \ref{fractionsim} also shows that the simulated sample has a similar relation between the stellar mass fraction and the total sub-halo mass with respect to our model. We note that the sub-haloes extracted from the simulated catalogue have a slightly higher (around $3 \times 10^{10} \, M_\odot$) lower limit for the total mass compared to those extracted from the SL model adopting the same criterion. This seems to suggest that at a fixed velocity dispersion simulated sub-haloes have, on average, a higher total mass. In the next subsection we address this issue in light of the recent studies that have focused on it.  
   
   \subsection{Compactness of the cluster sub-haloes}
   
   Significant differences between the sub-halo compactness obtained from lensing models and cosmological simulations have recently been reported by M20. They considered $11$ SL models of massive galaxy clusters, including the model of AS1063 by B19, and a set of $25$ clusters of similar mass and redshift from the cosmological simulations presented in \citet{planelles14} and \citet{rasia15}, which incorporate gas cooling, star formation, and energy feedback from supernovae and accreting supermassive black holes. They found that the probability of producing galaxy-scale SL events is for observed clusters around an order of magnitude larger than it is for simulated clusters.
   To interpret this discrepancy, they considered the relation between the mass of the sub-haloes in the model and their maximum circular velocity, defined as $v_\mathrm{max}= \mathrm{max}\sqrt{\frac{GM}{r}}$ (for an isothermal model, $v_\mathrm{max} = \sqrt{2} \sigma_0$). M20 found that, at fixed total sub-halo mass, the value of the maximum circular velocity of cluster sub-haloes is larger than what cosmological simulations predict. This seems to indicate that simulated cluster members are, on average, less compact than observed.
   
   Some other recent results, using the higher-resolution simulations from the Hydrangea/C-EAGLE suite \citep[][]{bahe17,barnes17}, do not claim the same large discrepancy in terms of galaxy scale lensing events and of sub-halo compactness \citep[][hereafter B21]{robertson21,bahe21}. Studying simulated clusters, B21 reported a bi-modal $v_\mathrm{max}(M)$ function, with a lower sequence of sub-haloes with a baryon fraction smaller than $0.1$ dominating for $M<10^{11} \, M_{\odot}$, and a second sequence for higher sub-halo mass values. This last branch yields results consistent with those of the lensing models presented in M20. %Finally, they have found that taking into account the effect of baryons leads to higher $v_\mathrm{max}$ values, at fixed total mass.
   
\begin{figure}
  \centering
    \includegraphics[width=9cm]{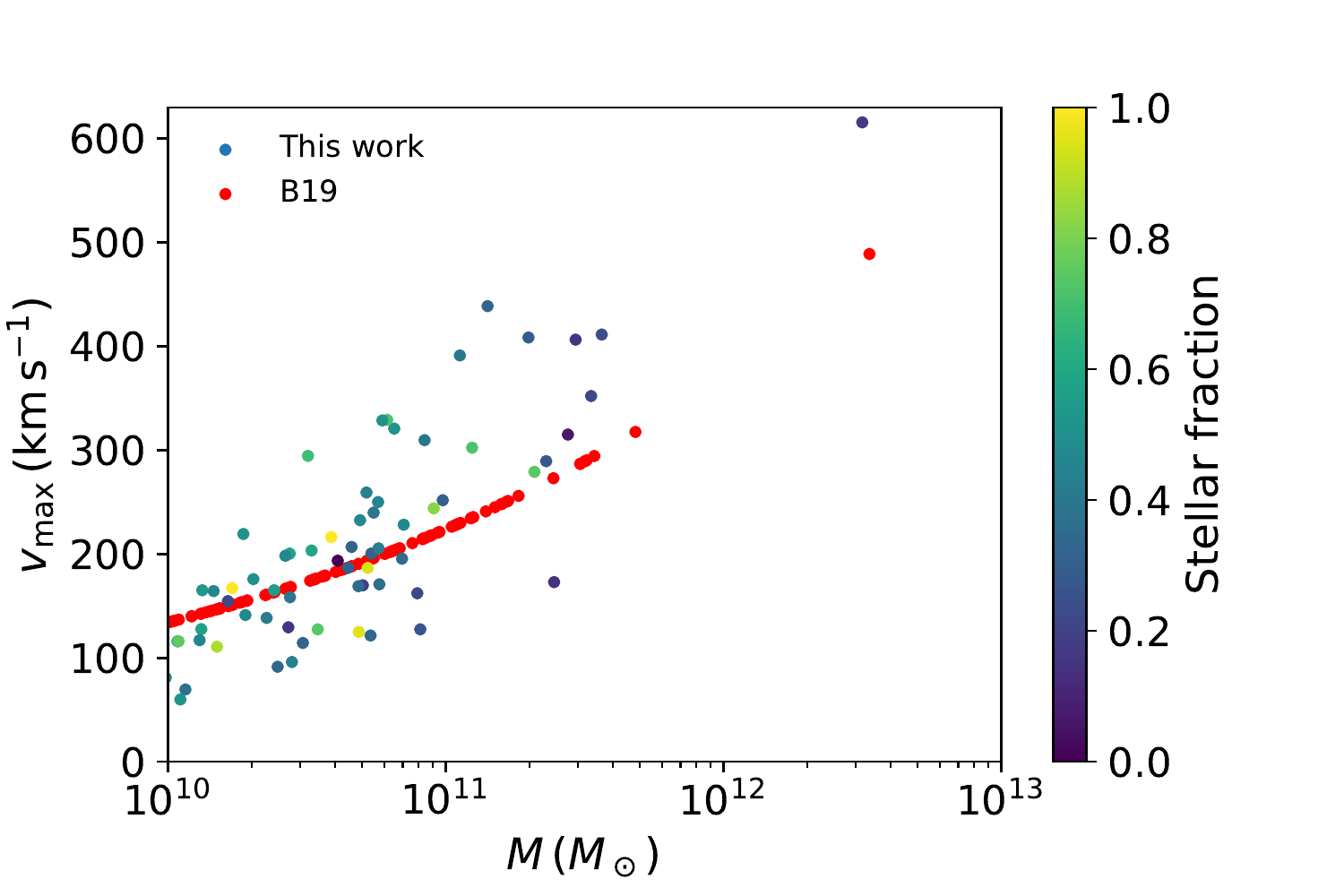}
  \caption{Maximum circular velocity of the cluster members as a function of their total mass. The values predicted by the model presented in B19 are shown in red. Our values are colour-coded according to the predicted stellar over total mass fraction. Only members with $M>10^{10} \, M_{\odot}$ are shown.}
  \label{bahe}
  \end{figure}
  \begin{figure}
  \centering
    \includegraphics[width=9cm]{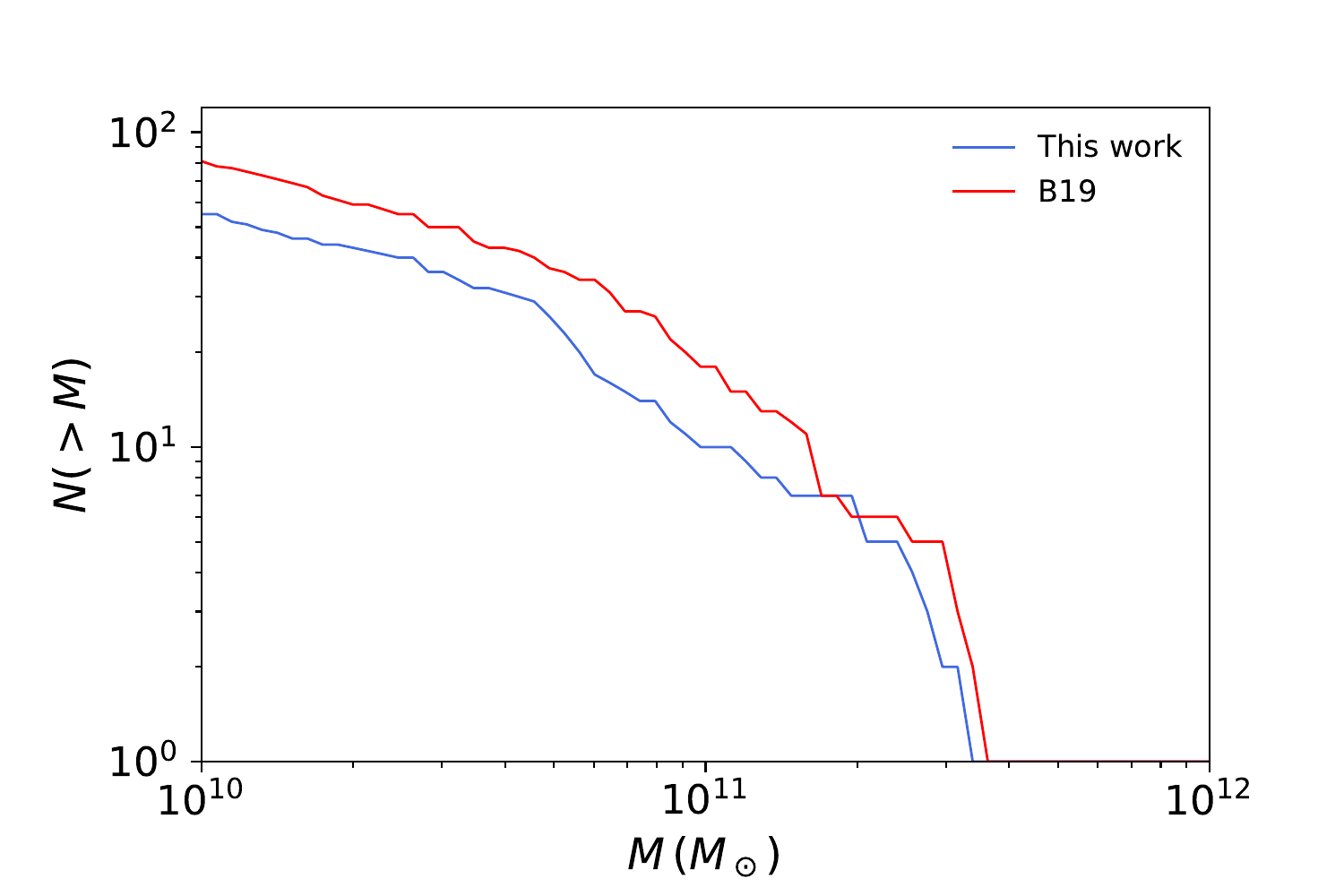}
  \caption{Cumulative sub-halo mass function for the members within a projected radius of $0.15 r_\mathrm{vir}$ predicted by our model (in blue) by that presented in B19 (in red).}
  \label{bahe2}
  \end{figure}
   
   All SL models included in M20 used two optimised power-law scaling relations to model the truncation radius and the velocity dispersion of the cluster members from their luminosity \citep[B19,][]{caminha19}. As all cluster models represent sub-haloes with isothermal profiles, this also leads to a fixed power-law relation between the maximum circular velocity (which is proportional to the velocity dispersion) of each sub-halo and its mass. On the other hand, our new method to assign mass to sub-haloes again avoids a fixed power law between the two quantities, allowing for the inclusion of some scatter. In Fig. \ref{bahe}, we show the $v_\mathrm{max}$($M$) relation obtained from our model for all the members with $M>10^{10} \, M_\odot$ compared with that from B19: the two models agree in the entire mass range considered. 
   
   Comparing our results with those shown in Fig. 3 from B21, we notice that our relation matches fairly well with the higher-mass branch of the Hydrangea hydrodynamic simulations, both in terms of circular velocity and of the stellar fraction of the cluster members at a fixed total mass. However, we note that the high-mass branch that agrees with our model in terms of maximum circular velocity is not found, for instance, in the Illustris-TNG300s simulation \citep[][]{marinacci18,naiman18,nelson18,pillepich18,springel18}, which have a comparable resolution, as shown in Fig. 4 from B21. Furthermore, \citet{ragone18} also found that the Hydrangea simulations produce galaxies with a very high total mass that may not agree with observations. On the other hand, as already visible from Fig. \ref{bahe}, we do not find the large number of sub-haloes with $M<10^{11} \, M_\odot$, $v_\mathrm{max}<100$ $\mathrm{km \, s^{-1}}$, and a low stellar fraction predicted by the lower sequence by B21. As a consequence, our model predicts, on average, higher maximum circular velocities with respect to simulations for sub-haloes with mass between $10^{10}$ and $10^{11} \, M_\odot$. In this total mass range, we find just a few members with maximum circular velocity lower than $100 \, \mathrm{km \, s^{-1}}$, while in B21 most of the galaxies lie below this threshold. This seems to confirm the discrepancy reported by M20 in terms of galaxy-scale SL events, as most of these events are caused by haloes in this mass range, at least for AS1063 and the other clusters with a similar mass. We finally note that in B21 the values of the baryon fractions are computed within a radius of $30 \, \mathrm{kpc}$, while in our work they are the ratio between the total stellar mass and the total mass of each sub-halo. To compare the values obtained in the two cases, one must keep in mind that our model predicts that, for the high-mass cluster members, a median value of $94\%$ of the total mass is enclosed within a sphere with a radius of $30 \, \mathrm{kpc}$.
   
   The comparative lack of low-mass members in our model is reflected by the cumulative sub-halo mass function within a projected radius of $0.15 r_\mathrm{vir}$, shown in Fig. \ref{bahe2}: our model agrees with the results presented in B19, which in turn found a lower cumulative mass function compared to that of the simulations, particularly considering sub-haloes with $M<10^{11} \, M_\odot$ and $v_\mathrm{max}<100$ $\mathrm{km \, s^{-1}}$, as reported by B21 in Fig. 1. In the same mass range, our cumulative sub-halo mass function has a slope similar to that by B19, significantly shallower than that from the cosmological simulations presented in B21. In the radial range considered, our sample of photometric cluster members is highly complete down to $m_\mathrm{F160W}=24$. 
   
\section{Conclusions}\label{s8}
   
   In this article we have presented an improved SL model of the HFF galaxy cluster AS1063 thanks to new measurements of the structural parameters, namely magnitude, effective radius, and S\'ersic index, of the cluster members. 
   We have also exploited the values of the central stellar velocity dispersion of several luminous members, obtained from MUSE integral-field spectroscopy, to calibrate the FP relation for the cluster.
   
   We used the FP to estimate the value of the central velocity dispersion of each cluster member from its observed magnitude and effective radius, and we chose to consider a proportionality relation between the values of the effective radii, $R_e$, and truncation radii, $r_t$. This allowed us to assign a value of total mass to every cluster member that does not depend only on the total luminosity of a galaxy. The main results and conclusions of our analysis are summarised as follows: 
   
   \begin{enumerate}
      \item We find that the lowest value of the rms difference between the observed and model-predicted positions of the 55 considered multiple images (from 20 background sources) is obtained when $r_t=2.3R_e$ for all cluster members. 
      \item Our method allows us to determine a relation between the total mass and velocity dispersion of the cluster members that is more realistic than the simple power law used in B19. Specifically, the new relation has some visible scatter and is less steep than that recently found in B19.
      \item %Having determined more precisely the mass of the cluster members from observations, we are able to reduce the degeneracy between the parameters of the diffuse, cluster-scale, DM haloes. This emerges clearly considering 
      A more accurate estimate of the total mass of the cluster members allows for a reduction in the scatter on the determination of some parameters of the diffuse component. In particular, the statistical uncertainty on the value of the core radius of the main DM halo decreases by more than $30\%$ compared to previous studies.
      \item We confirm that modelling the main DM haloes of massive galaxy clusters with isothermal mass density profiles requires the presence of significantly large core radii. The value of the core radius of the main DM halo of AS1063 is $86 \pm 2 \, \mathrm{kpc}$.
      \item From our model and from the sampling of the posterior probability distribution of the parameters, we obtain the cumulative projected mass profiles of the various cluster components. Thanks to Chandra X-ray observations and new measurements of the stellar mass values of the cluster members, we can disentangle the hot gas and stellar mass profiles. At a distance from the cluster centre of $350 \, \mathrm{kpc}$, the cumulative projected stellar, hot-gas, and baryonic fractions are, respectively, $0.6\%$,  $14.1\%$, and $14.7\%$; the last value is close to the cosmological value found by \citet{planck20}.
      \item For the high-mass members included in our model, the median value of the stellar over total mass fraction is $0.47$, with a standard deviation of $0.34$. Considering, instead, the ratio between the stellar and total mass projected within the effective radius, we find good agreement with the relation obtained by \citet{grillo10} on a wide sample of SDSS early-type galaxies.
      \item We used the SHMR from \citet{girelli20}, based on the $\mathrm{\Lambda}$CDM DUSTGRAIN-\textit{pathfinder} $N$-body simulation, to obtain the halo mass values of the cluster members from their measured stellar mass values. We find that the SHMR predicts a median value of the stellar fraction of the high-mass cluster members that is smaller by almost an order of magnitude with respect to our model.
      \item We compared the high-mass cluster members with a catalogue of simulated sub-haloes. We extracted the sub-haloes from the projected cores of $18$ clusters from recent cosmological hydrodynamical simulations \citep[][]{planelles14,rasia15}, imposing the same selection criterion on their velocity dispersion. We report a good agreement between the values of the stellar over total mass fraction of the two samples.
      \item We studied the relation between the maximum circular velocity and total mass of the sub-haloes, which is an indication of their compactness. Comparing our results with those obtained by B21 from the simulated clusters of the Hydrangea/C-EAGLE suite, we obtain consistent results for sub-haloes with total mass greater than $10^{11} \, M_\odot$. On the contrary, we do not observe the large number of sub-haloes with mass below $10^{11} \, M_\odot$ and low baryon fraction predicted by B21. In this sub-halo mass range, our relation predicts, on average, higher $v_\mathrm{max}$ at a fixed total mass, in agreement with B19 and M20. This corresponds to haloes that are more compact than those in the hydrodynamical simulations. As a consequence, in the same total mass range our cumulative sub-halo mass function has a shallower slope compared to what was reported by B21 and very similar to that from B19 and M20.   
   \end{enumerate}
   
   Similar analyses can be performed on other massive clusters, for which MUSE data are available, to test the robustness of the results obtained with the new methodology presented here.

\begin{acknowledgements}
We thank the anonymous referee for some useful suggestions that helped to improve the paper. We  acknowledge  financial  support  by  PRIN-MIUR 2017WSCC32 "Zooming into dark matter and proto-galaxies with massive lens-ing  clusters" (P.I.: P. Rosati),  INAF  “main-stream”  1.05.01.86.20 (P.I.:  M.  Nonino)  and  INAF  1.05.01.86.31 (P.I.: E. Vanzella). GG thanks Ana Acebron Mu{\~n}oz and Andrea Bolamperti for their help and suggestions. PB acknowledges financial support from ASI through the agreement ASI-INAF n. 2018-29-HH.0. MM acknowledges support from ASI through the grants ASI-INAF n. 2018-23-HH.0 and ASI-INAF n.2017-14-H.0. GBC acknowledges the Max Planck Society for financial support through the Max Planck Research Group for S. H. Suyu and the academic support from the German Centre for Cosmological Lensing.
\end{acknowledgements}

\bibliographystyle{aa} % style aa.bst
\bibliography{bibl.bib} % your references Yourfile.bib

\end{document}